\documentclass[a4paper,fleqn,usenatbib]{mnras}
\usepackage{newtxtext,newtxmath}
\usepackage[T1]{fontenc}
\usepackage{ae,aecompl}

\usepackage{graphicx}	% Including figure files
\usepackage{amsmath}	% Advanced maths commands
\usepackage{amssymb}	% Extra maths symbols

\title[Polycyclic Aromatic Hydrocarbon Features with Galaxy Merger]{A Relationship of Polycyclic Aromatic Hydrocarbon Features with Galaxy Merger in Star-forming Galaxies at $z<0.2$}

\author[K. L. Murata et al.]{
Katsuhiro L. Murata,$^{1}$\thanks{E-mail: katsuhirolmurata@gmail.com, murata@u.phys.nagoya-u.ac.jp} 
Rika Yamada,$^{1}$
Shinki Oyabu,$^{1}$\thanks{E-mail: oyabu@u.phys.nagoya-u.ac.jp}
Hidehiro Kaneda,$^{1}$ \newauthor
Daisuke Ishihara,$^{1}$
Mitsuyoshi Yamagishi,$^{2}$
Takuma Kokusho,$^{1}$
Tsutomu T. Takeuchi$^{1}$
\\
$^{1}$Department of Particle and Astrophysical Science, Nagoya University, Furo-cho, Chikusa-ku, Nagoya, 464-8602, Aichi, Japan\\
$^{2}$Institute of Space and Astronautical Science, Japan Aerospace Exploration Agency, Sagamihara, 229-8510, Kanagawa, Japan
}

\date{Accepted 2017 July 22. Received 2017 July 6; in original form 2016 October 13.}

\pubyear{2016}

\begin{document}
\label{firstpage}
\pagerange{\pageref{firstpage}--\pageref{lastpage}}
\maketitle

\begin{abstract}
Using the AKARI, Wide-field Infrared Survey Explorer (WISE), Infrared Astronomical Satellite (IRAS), Sloan Digital Sky Survey (SDSS) and Hubble Space Telescope (HST) data, we investigated the relation of polycyclic aromatic hydrocarbon (PAH) mass ($M_{\rm PAH}$), very small grain mass ($M_{\rm VSG}$), big grain mass ($M_{\rm BG}$) and stellar mass ($M_{\rm star}$) with galaxy merger for 55 star-forming galaxies at redshift $z<0.2$. 
Using the SDSS image at $z<0.1$ and the HST image at $z>0.1$, we divided the galaxies into merger galaxies and non-merger galaxies with the morphological parameter asymmetry $A$, and quantified merging stages of galaxies based on the morphological indicators, the second-order momentum of the brightest 20$\%$ region $M_{20}$ and the Gini coefficient. 
We find that $M_{\rm PAH}/M_{\rm BG}$ of merger galaxies tend to be lower than that of non-merger galaxies and there are no systematic differences of $M_{\rm VSG}/M_{\rm BG}$ and $M_{\rm BG}/M_{\rm star}$ between merger galaxies and non-merger galaxies. We find that galaxies with very low $M_{\rm PAH}/M_{\rm BG}$ seem to be merger galaxies at late stages. These results suggest that PAHs are partly destroyed at late stages of merging processes. Furthermore, we investigated $M_{\rm PAH}/M_{\rm BG}$ variations in radiation field intensity strength $G_0$ and the emission line ratio of $[\ion{O}{i}]\lambda6300/{\rm H}\alpha$ which is a shock tracer for merger galaxies and find that $M_{\rm PAH}/M_{\rm BG}$ decreases with increasing both $G_0$ and $[\ion{O}{i}]/{\rm H}\alpha$. PAH destruction is likely to be caused by two processes; strong radiation fields and large-scale shocks during merging processes of galaxies. 
\end{abstract}


%%%%%%%%%%%%%%%%%%%%%%%%%%%%%%%%%%%%%%%%%%%%%%%%%%

%%%%%%%%%%%%%%%%% BODY OF PAPER %%%%%%%%%%%%%%%%%%

\section{Introduction}
Polycyclic aromatic hydrocarbons (PAHs), large planar molecules with $50-1000$ carbon atoms, are a ubiquitous component of galaxies in the Universe. PAHs are quite easily destroyed by various physical processes, which include collisions with high-velocity electrons and ions in shock regions induced by supernovae (SNe) and hot plasma, strong UV radiation of massive young stars, and hard UV and soft X-ray radiation of active galactic nuclei (AGN). Several studies found that PAH emission in supernova remnants (SNRs) is significantly suppressed relative to dust emission (e.g., \citealt{Ishihara:2010fi}), which can be explained by low abundance of PAHs relative to dust through selective destruction of PAH by SN-induced shocks (e.g., \citealt{Tielens:2008fx}). In low-metallicity starburst galaxies, supernova-induced shocks are considered to be a more dominant process for destruction of PAHs than UV radiation from massive stars \citep{OHalloran:2006eu}, although the latter is crucial for PAH destruction in low-metallicity blue compact dwarf galaxy \citep{Plante:2002jk}.

Merging of galaxies is considered to induce large-scale strong shocks in the galaxies; PAH destruction on a galaxy scale is expected to occur during this merging processes. Recent integral field spectroscopic observations found that some nearby galaxies possess extended ionized gas with high velocity dispersions, which is interpreted as galaxy-wide shocks with speeds of $150-500$~km/s induced by merging of galaxies (\citealt{MonrealIbero:2006fi,MonrealIbero:2010gz}; \citealt{Rich:2011is,Rich:2014ib,Rich:2015kf}). In the hierarchical picture of a galaxy assembly, galaxies have evolved through merging processes. Since the merging of galaxies is common especially at higher redshifts (e.g., \citealt{Lotz:2011bu}), it is crucial to explore the PAH destruction during a merging process in order to understand the history of PAH enrichment for various types of galaxies in the Universe.

The number of systematic studies about PAH destruction through merger-induced shocks is, however, rather limited, because a large sample of star-forming galaxies during merging processes is required. Since some galaxies in merging processes have AGNs which also destroy their PAHs by hard UV and soft X-ray radiations, it is necessary to divide galaxies into pure star-forming galaxies and galaxies with AGNs in order to investigate the PAH destruction by merger-induced shocks.
In our previous study of \citet{Yamada:2013ca}, we constructed the clean sample of star-forming galaxies based on the PAH $3.3\,\micron$ feature equivalent width and near-infrared (NIR) slope using spectroscopic observations by the AKARI satellite \citep{Murakami:2007bs} and investigated their $3.3\,\micron$ PAH luminosity ($L_{\rm PAH3.3}$) and total infrared luminosity ($L_{\rm TIR}$). We find that $L_{\rm PAH3.3}/L_{\rm TIR}$ systematically decreases with $L_{\rm TIR}$, which is interpreted as decreases in the amount of PAHs relative to dust grains with $L_{\rm TIR}$. Combining the decrease of $L_{\rm PAH3.3}/L_{\rm TIR}$ and the previous observations that a merger fraction increases with increasing $L_{\rm TIR}$, \citet{Yamada:2013ca} suggested that PAHs are destroyed via shocks induced by merging processes of galaxies.

In this paper, we aim to investigate the relation of PAH features ($L_{\rm PAH}$) with galaxy merger for star-forming galaxies, using optical images of the Sloan Digital Sky Survey (SDSS) and Hubble Space Telescope (HST). Optical images with high spatial resolution allow us to identify galaxy merger remnants and thus securely classify galaxies into merger galaxies and non-merger galaxies. 

\section{Sample of Star-forming Galaxies}
Our sample consists of 57 star-forming galaxies in the sample of \citet{Yamada:2013ca}, out of which 48 galaxies were observed with SDSS DR9 \citep{Ahn:2012ih} at redshift $z<0.1$ and the other 9 galaxies were observed with HST at redshift $0.1<z<0.2$. The sample of \citet{Yamada:2013ca} was based on the two AKARI mission programs, Mid-infrared Search for Active Galactic Nuclei (MSAGN: \citealt{Oyabu:2011kh}) and Evolution of ULIRGs and AGNs (AGNUL: \citealt{Imanishi:2008wj,Imanishi:2010jw}). 
These two programs observed galaxy spectra at $2.5-5\,\micron$ wavelengths with a spectral resolution of $R\sim120$ \citep{Ohyama:2007gm} using the AKARI/IRC (InfraRed Camera; \citealt{Onaka:2007pj}), which enables to detect the PAH $3.3\,\micron$ feature and to measure the NIR continuum slope. 
Galaxies with either the equivalent width of the $3.3\,\micron$ feature $<40$~nm (\citealt{Moorwood:1986wa}; \citealt{Imanishi:2000ev}; \citealt{Imanishi:2008wj}) or NIR continuum slope $\Gamma>1$ ($f_\nu\sim\lambda^\Gamma$) (\citealt{Risaliti:2006bm}; \citealt{Imanishi:2010jw}) were excluded as those likely contaminated by AGN activity. 

We constructed a sample of 48 galaxies with the SDSS $g$-band images at $z<0.1$ from the sample of \citet{Yamada:2013ca}. The redshift boundary is determined to avoid the uncertainties of galaxy morphology estimation due to lower physical spatial resolution toward higher redshifts, which is important for secure morphological classification of our sample into merger galaxies and non-merger galaxies. 
The pixel scale and typical full width at half-maximum (FWHM) size of Point Spread Function (PSF) are $0.396$ arcsec and $\sim1.5$ arcsec, respectively. The PSF size corresponds to the physical resolution of 2.8 kpc at $z=0.1$ which is sufficient to resolve merger features. 

% HST 
For the galaxies of \citet{Yamada:2013ca} at $z>0.1$, we searched the optical images with high spatial resolutions of the HST in the Hubble Legacy Archive{\footnote{ http://hla.stsci.edu/} and constructed a sample of 9 galaxies with the HST F814W band images. The HST galaxies are at spectroscopic redshift $z=0.108-0.168$. The pixel scale and the PSF FWHM are 0.10 arcsec and $\lesssim0.2$ arcsec, respectively. The PSF size corresponds to the physical resolution of $\lesssim0.6$ kpc at $z=0.168$. 

We searched for companion galaxies of the 57 galaxies. 
For the SDSS sample, we defined sources with `galaxy' class in the SDSS catalog within $150$~kpc from the primary galaxy positions and within relative velocity along line-of-sight of $500$~km/s from the primary galaxies as companion galaxies. The relative velocities were estimated from the SDSS spectroscopic redshifts. 
We determined the boundaries used in the above definition of companion galaxies according to the merger simulation by \citet{Lotz:2010hf}. 
For the HST sample, we searched for companion galaxies in the HST image by visual inspection.
As a result, out of all the 57 galaxies in our sample, 21 galaxies are found to have a companion galaxy. 

\begin{figure*}
 \begin{minipage}{0.45\hsize}
  \begin{center}
   \includegraphics[width=75mm]{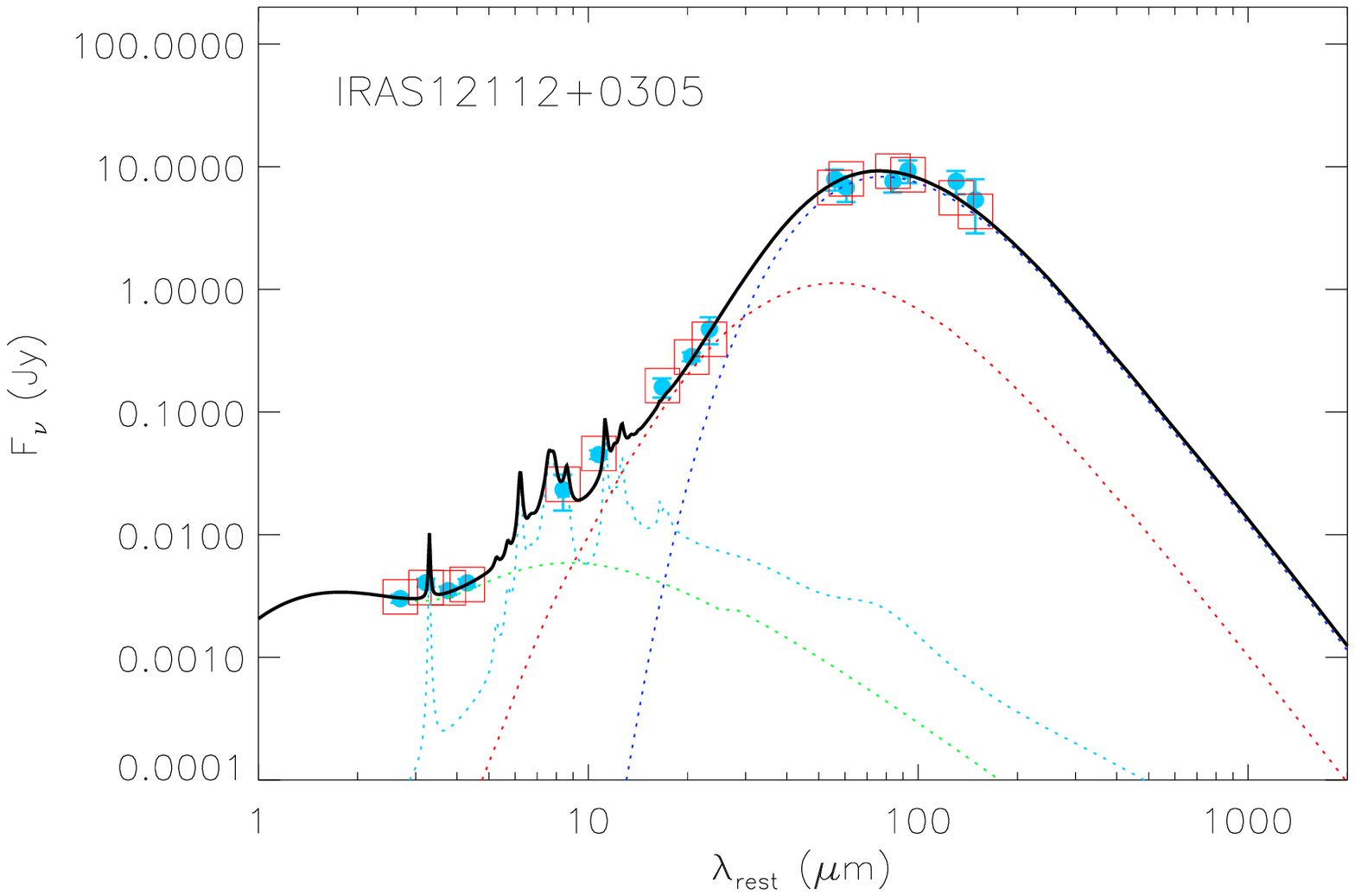}
  \end{center}
 \end{minipage}
 \begin{minipage}{0.45\hsize}
  \begin{center}
   \includegraphics[width=75mm]{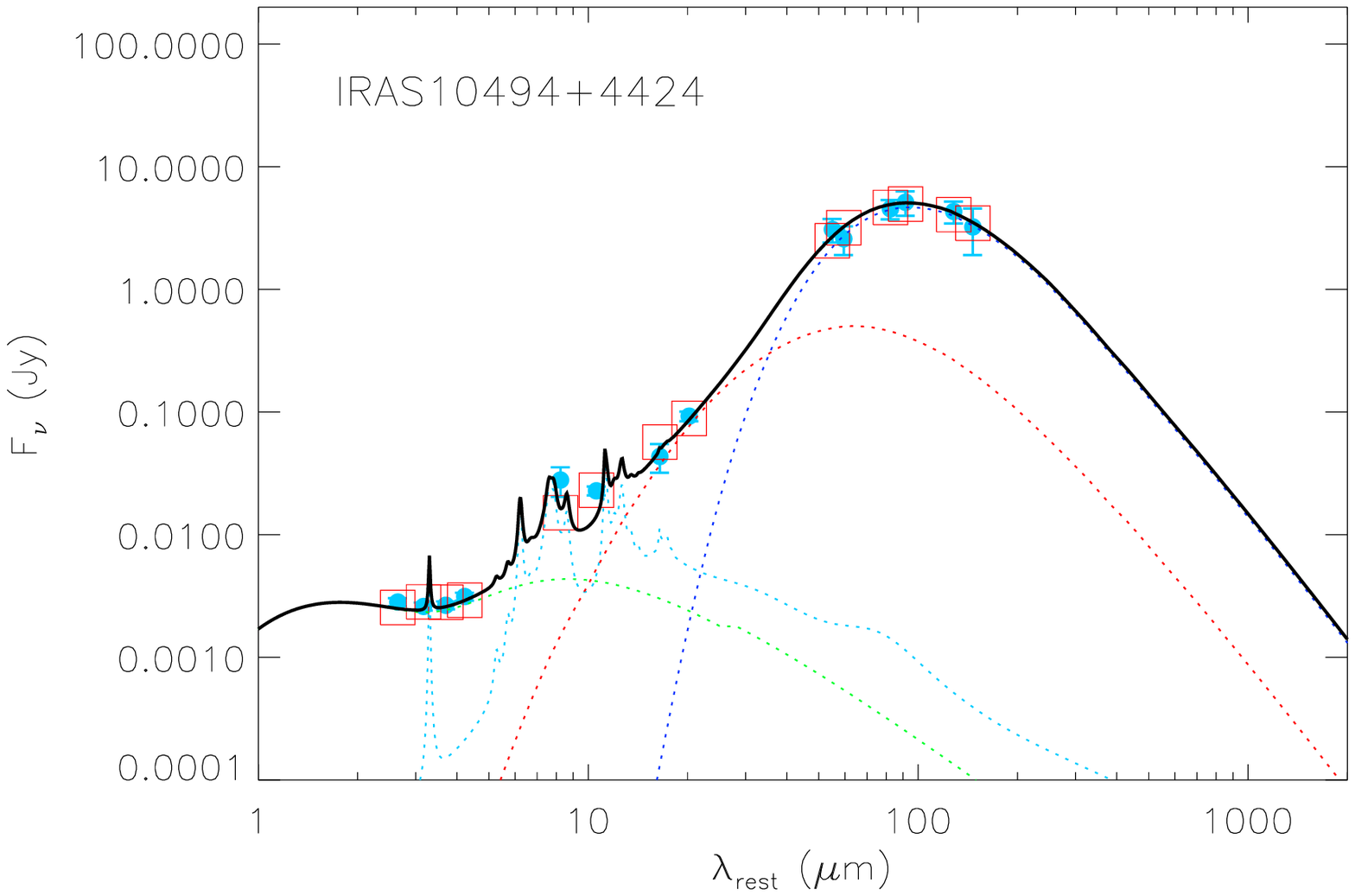}
  \end{center}
 \end{minipage}
%%%%%%%%%%%%%%%%%%%%%%%%%%%%%%%%%%%%%%%%%%%%%%%%%%%%%%
 \begin{minipage}{0.45\hsize}
  \begin{center}
   \includegraphics[width=75mm]{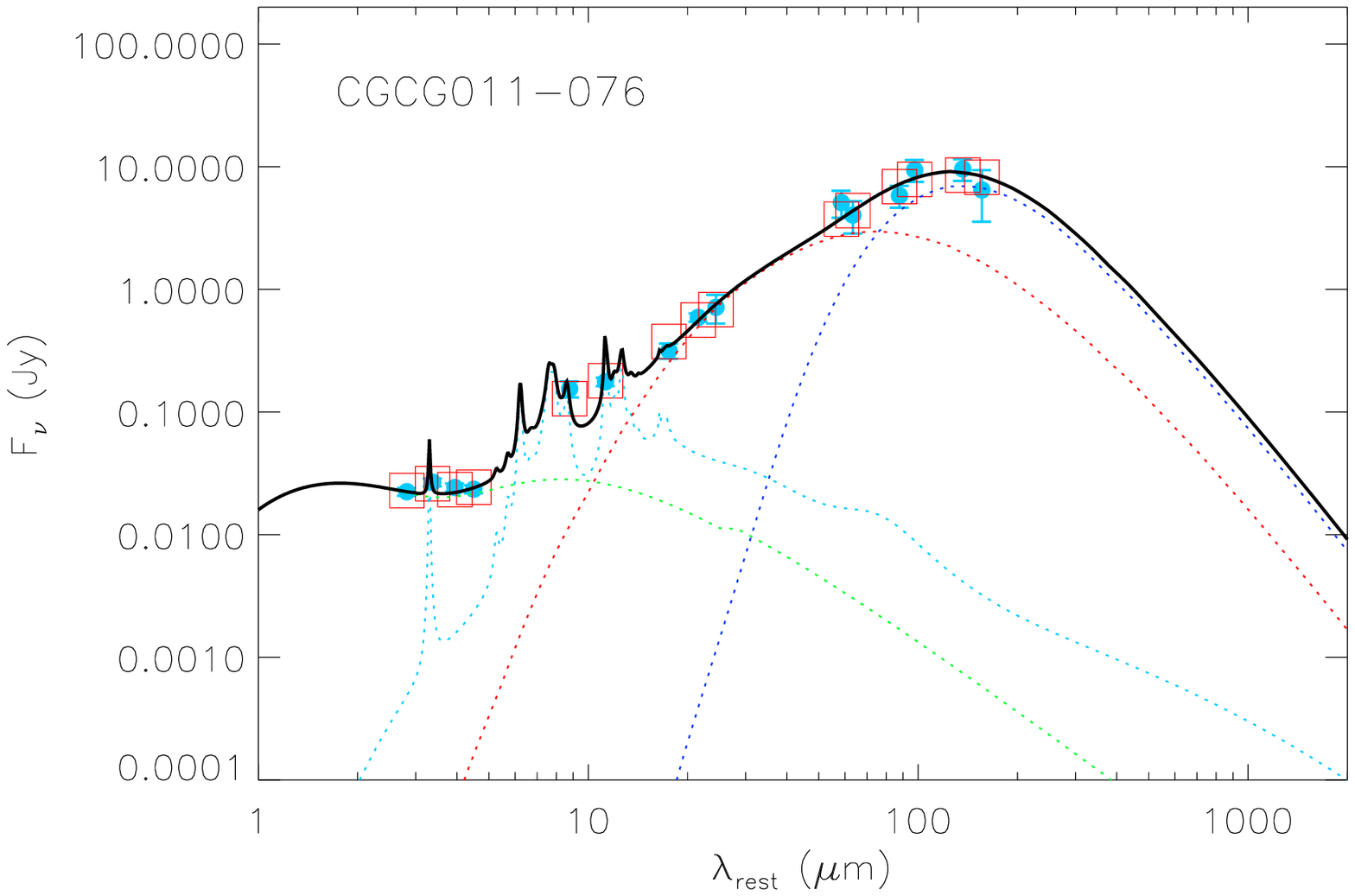}
  \end{center}
 \end{minipage}
 \begin{minipage}{0.45\hsize}
  \begin{center}
   \includegraphics[width=75mm]{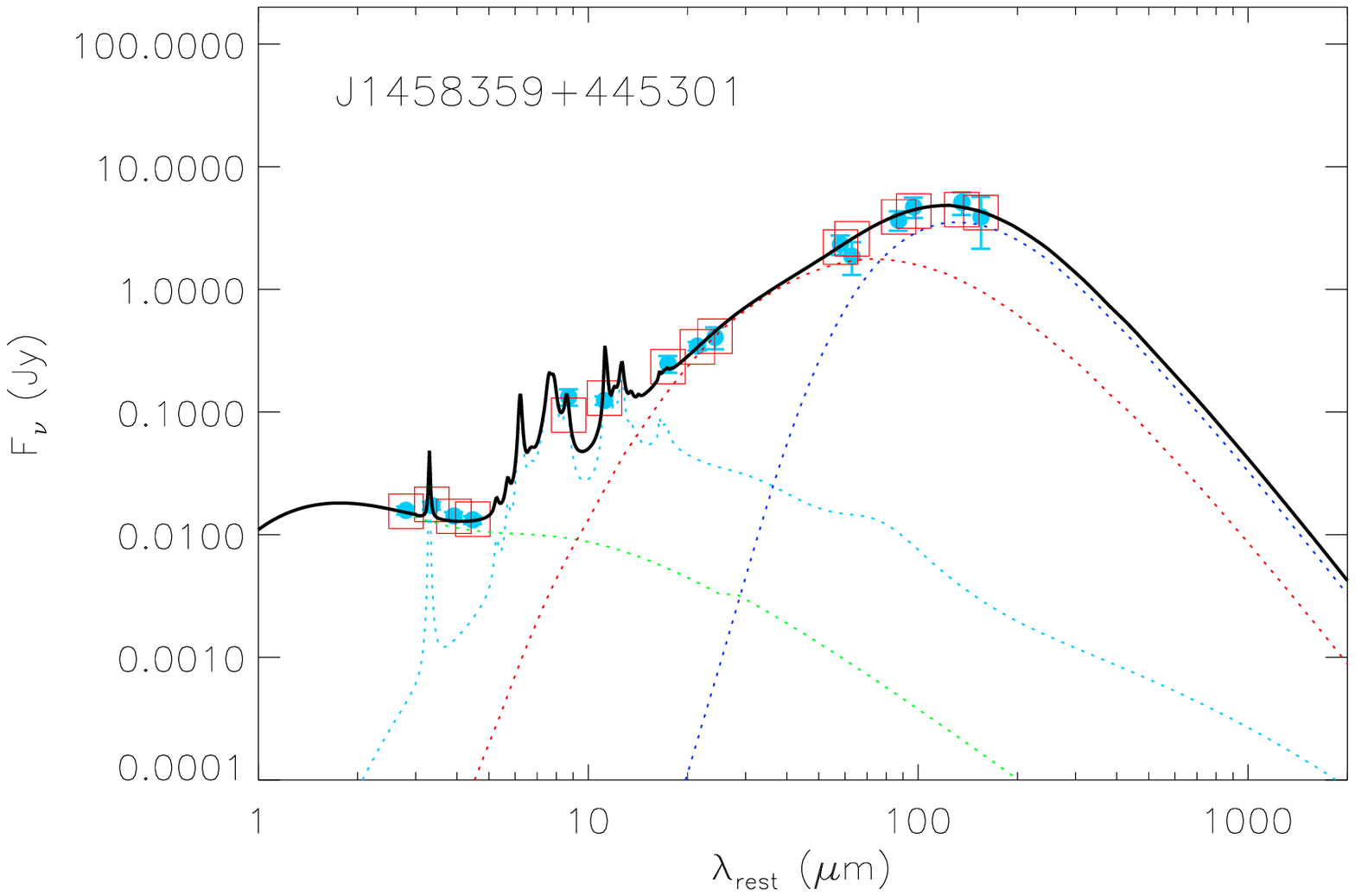}
  \end{center}
 \end{minipage}
%%%%%%%%%%%%%%%%%%%%%%%%%%%%%%%%%%%%%%%%%%%%%%%%%%%%%%
  \begin{minipage}{0.45\hsize}
  \begin{center}
   \includegraphics[width=75mm]{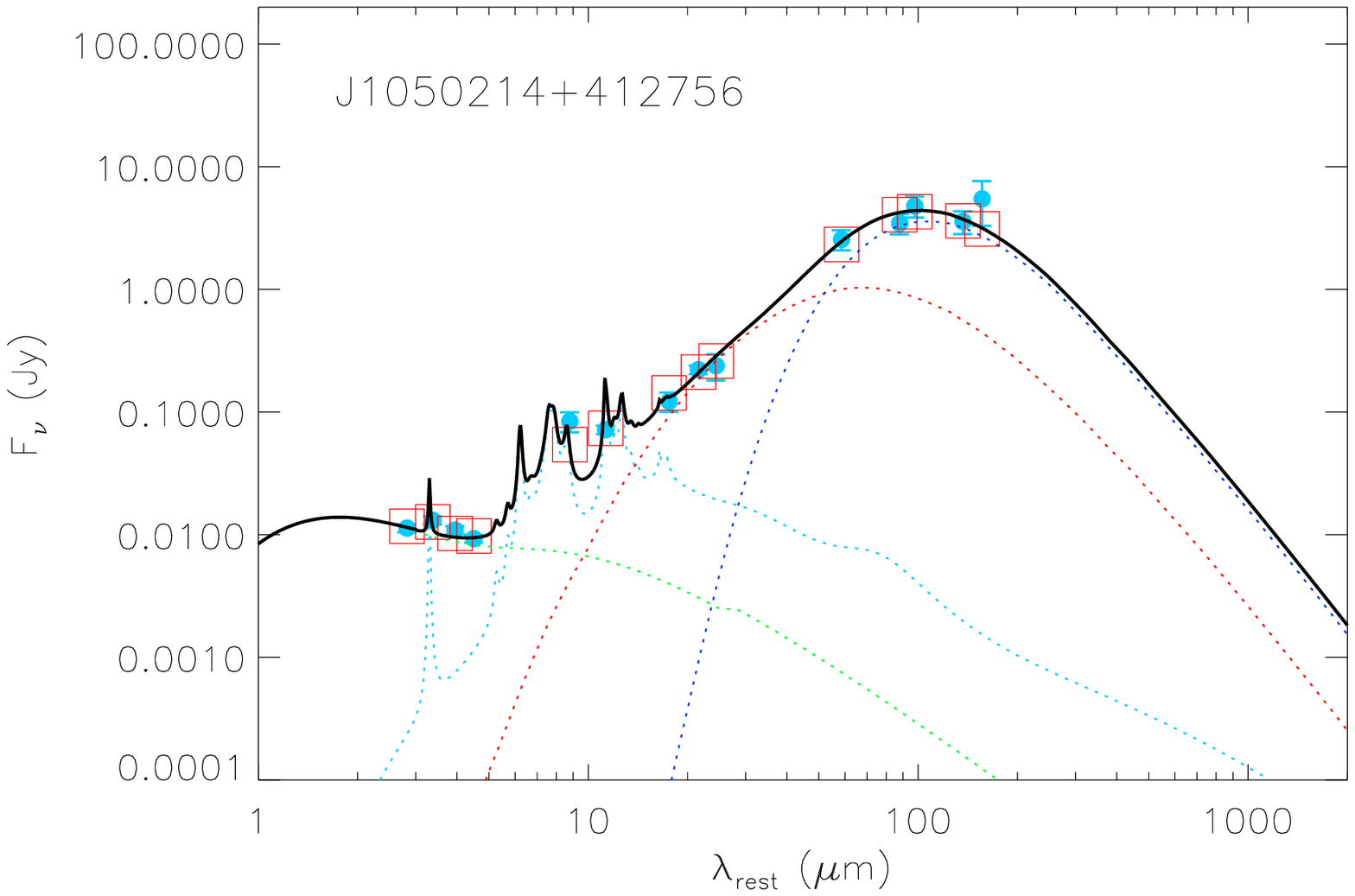}
  \end{center}
 \end{minipage}
 \begin{minipage}{0.45\hsize}
  \begin{center}
   \includegraphics[width=75mm]{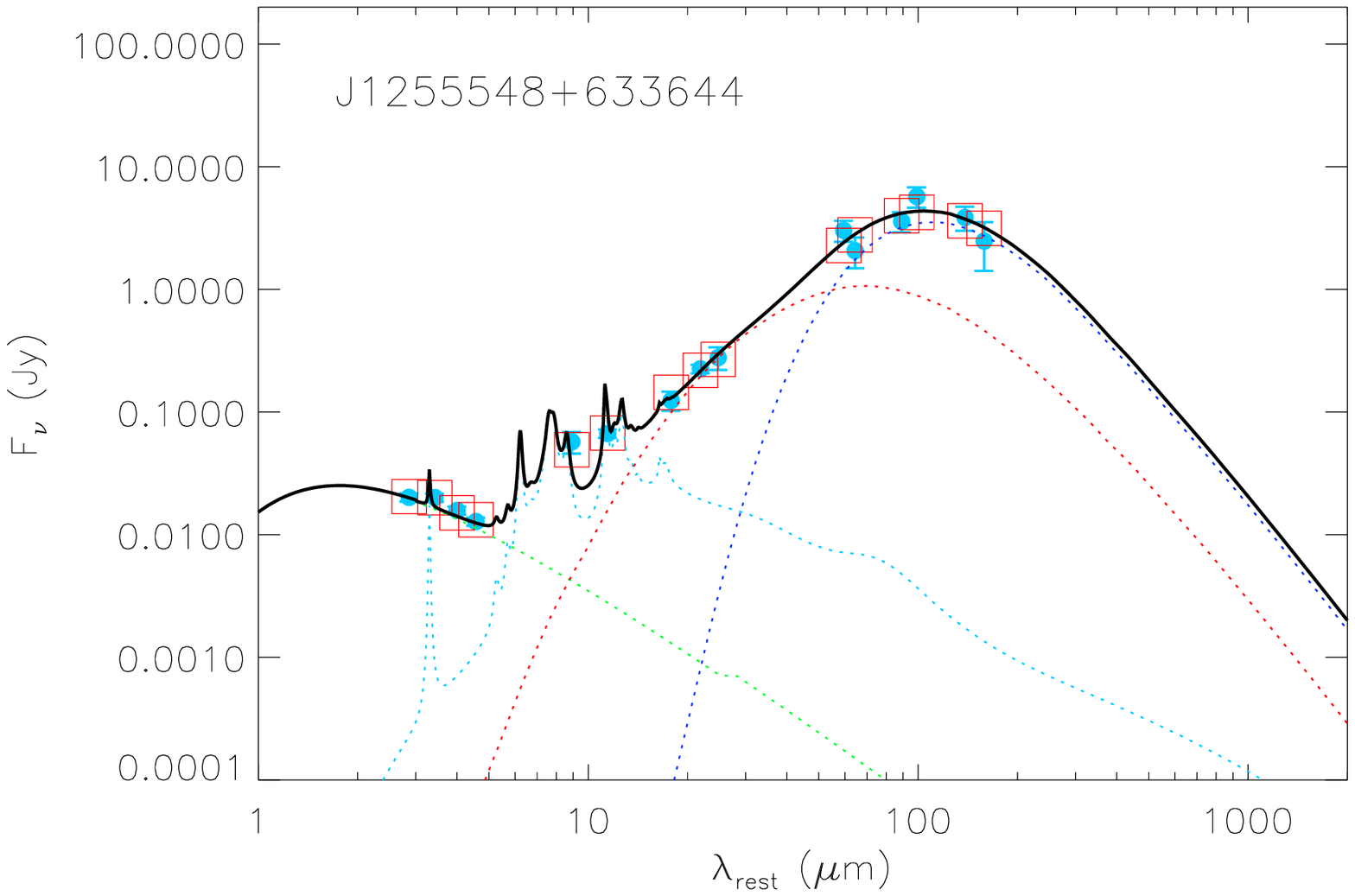}
  \end{center}
 \end{minipage}
\caption{
Examples of the observed SEDs. The model consists of the continua of 600\,K and 3000\,K blackbodies (green dotted line), PAHs (cyan dotted line), VSGs (red dotted line) and BGs (blue dotted line). The best-fit model spectrum and the model flux densities are plotted with the solid line and squares, respectively. The SEDs in the top, middle and bottom panels are taken from ULIRGs, LIRGs and IRGs, respectively.
}
\label{fig:sedfit}
\end{figure*}

\begin{figure*}
\begin{center}
\includegraphics[width=155mm]{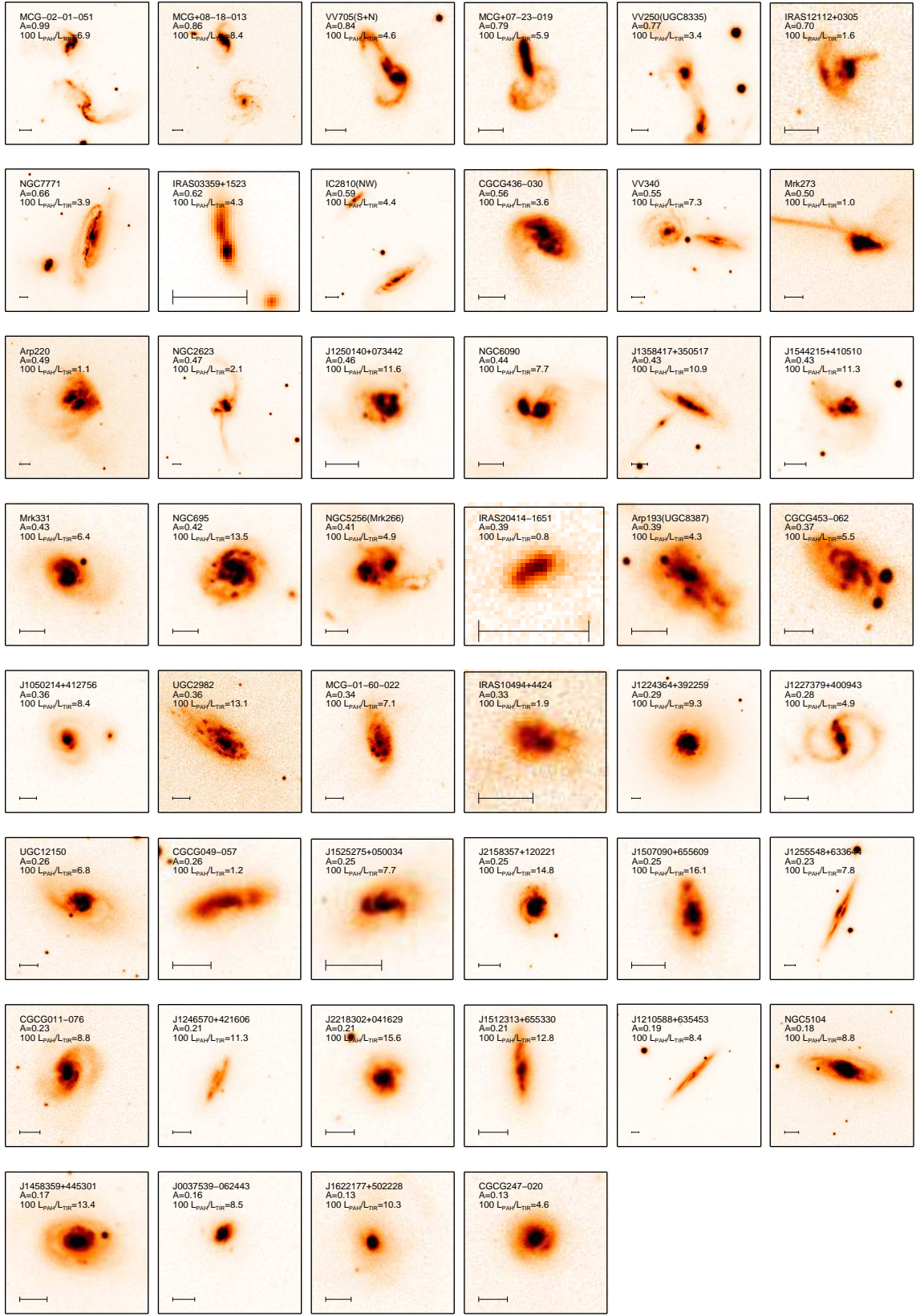}
\caption{
SDSS $g$-band images of the 46 SDSS galaxies. The images are ordered by descending asymmetry from the top left to the bottom right. The primary galaxy name, asymmetry value $A$ and $100\,L_{\rm PAH}/L_{\rm TIR}$ are given in each image. The scale bars are 10 arcsec for each image. 
}
\label{fig:thumb-sdss}
\end{center}
\end{figure*}

\begin{figure*}
\begin{center}
\includegraphics[width=60mm,angle=-90]{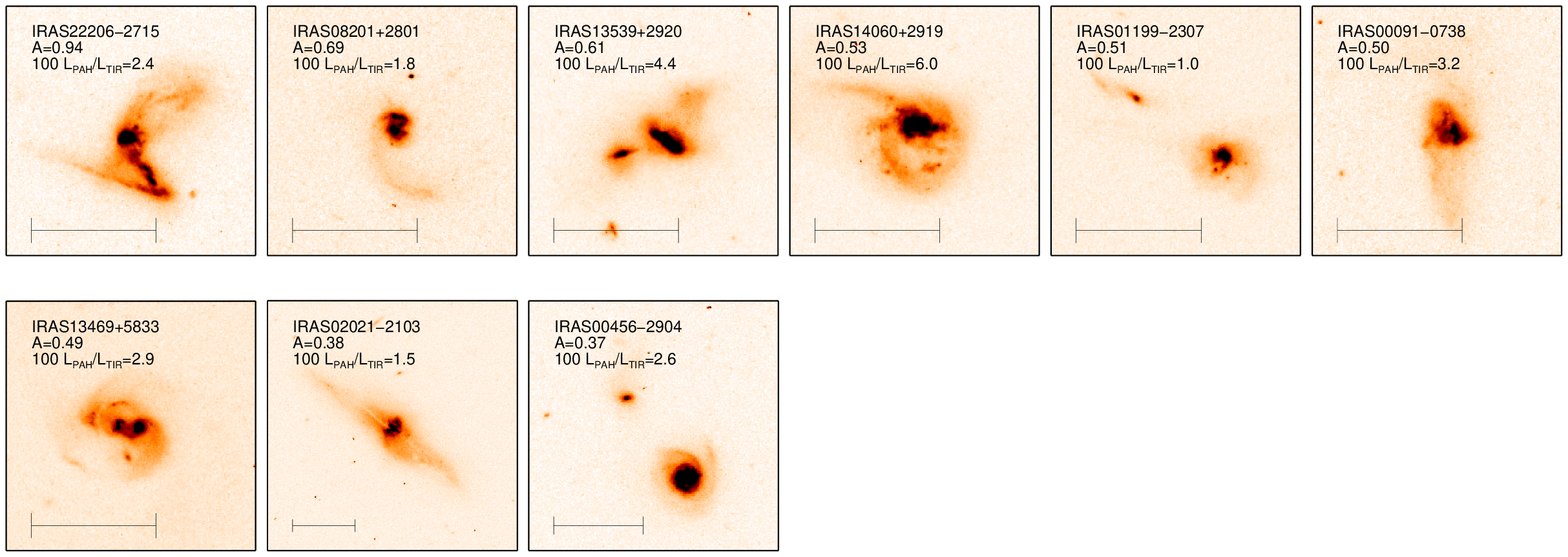}
\caption{
HST F814W-band images of the 9 HST galaxies. The scale bars are 10 arcsec for each image. 
}
\label{fig:thumb-hst}
\end{center}
\end{figure*}

\section{Data Analysis}
\subsection{Spectral Energy Distribution Fitting}
We estimated infrared properties of dust and PAH by a spectral energy distribution (SED) fitting technique using multi-band photometric data from NIR to FIR wavelengths.
We used the photometric data of the $9\,\micron$, $18\,\micron$, $65\,\micron$, $90\,\micron$, $140\,\micron$ and $160\,\micron$ bands of AKARI, the $12\,\micron$ and $22\,\micron$ bands of WISE (Wide-field Infrared Survey Explorer; \citealt{Wright:2010in}), the $25\,\micron$, $60\,\micron$ and $100\,\micron$ bands of IRAS (Infrared Astronomical Satellite; \citealt{Neugebauer:1984ch}), and the $2.5-5\,\micron$ spectrum of the AKARI/IRC for the SED fitting. The data of $2.5-5\,\micron$, $9\,\micron$ and $12\,\micron$ likely trace PAH emission features, while the other photometric data trace dust continuum emission. 

The $9\,\micron$ and $18\,\micron$ fluxes were measured from the AKARI mid-infrared all-sky diffuse maps (\citealt{Ishihara:2010gd}; Ishihara et al., in prep.) and the $65\,\micron$, $90\,\micron$, $140\,\micron$ and $160\,\micron$ fluxes were measured from the AKARI far-infrared all-sky diffuse maps \citep{Doi:2015cy}. The source fluxes were calculated by circular aperture photometry with background sky subtraction. The aperture sizes are flexibly changed to capture most of the galaxy light. 
Systematic uncertainties derived from the AKARI instruments and the photometric calibrations in the bands are included into the flux uncertainties with 15\%, 15\%, 20\%, 20\%, 30\% and 40\% for the $9\,\micron$, $18\,\micron$, $60\,\micron$, $90\,\micron$, $140\,\micron$ and $160\,\micron$ band respectively. 
The $12\,\micron$ and $22\,\micron$ fluxes were taken from the WISE All-Sky Catalog\footnote{http://wise2.ipac.caltech.edu/docs/release/allsky/}, and the $25\,\micron$, $60\,\micron$ and $100\,\micron$ fluxes were taken from IRAS Faint Source Catalog \citep{Moshir:1990vn}. 
The flux inconsistency between the IRAS and AKARI bands is included into the flux uncertainties, which is 15\% of the IRAS fluxes. The WISE photometry was calibrated using the standard stars, which is likely to introduce systematic offsets of the fluxes for galaxies with redder colors than the standard stars. Therefore systematic uncertainties are included into the flux uncertainties with 10\% for the $12\,\micron$ and $22\,\micron$ bands \citep{Wright:2010in}. 
The $2.5-5\,\micron$ spectra were reduced in \citet{Yamada:2013ca} using the AKARI/IRC pipeline. 
For the 21 galaxies which have a companion galaxy, we used the photometric flux as the sum of the fluxes of a primary galaxy and a companion galaxy because we could not separate these galaxies spatially with IRAS. 
If a primary galaxy and a companion galaxy are resolved spatially with AKARI and WISE, we evaluated each flux and added them. If not, we performed photometry as a single source with an appropriate aperture size. 

The SED model used in the fitting was generated using the {\tt DustEM} code \citep{Compiegne:2011jf}. 
The SED model is based on the DHGL (diffuse interstellar medium at high-galactic latitude) model of the {\tt DustEM}, which reproduces dust emission and extinction of the diffuse interstellar medium at high galactic latitudes, with some modifications to adjust for our data set. 
The DGHL model includes three dust components: amorphous silicate (aSil), hydrogenerated amorphous carbon (amC) and PAHs. The amC consists of different size components: large amC (LamC) and small amC (SamC). 
The PAHs are subdivided into neutral and ionized PAHs. The abundance ratio of LamC and aSil is assumed to be $1:5.4$ which reproduces dust emission and extinction for the GHDL of the Galaxy.
The abundance ratio of neutral and ionized PAHs is assumed to be $1.5:1$ corresponding to the ionization fraction of $0.4$ which is typical for average PAH size \citep{Li:2001aa}. 
In addition to the three dust components, two blackbody spectra with $3000$\,K as old stars which dominate NIR continuum in galaxies and $600$\,K which is hot dust temperature in infrared galaxies \citep{Oyabu:2011kh}} are included in the fitting model to reasonably fit the NIR spectra. 
The amplitudes of LamC and aSil, SamC, neutral PAH, ionized PAH, $3000$\,K and $600$\,K blackbodies are determined in the SED fitting. 

The fitting between model and observed SEDs was performed with $\chi^2$ minimization using {\tt DustEM wrapper}\footnote{http://dustemwrap.irap.omp.eu/}, the IDL wrapper program of the {\tt DustEM} code. 
The observed SEDs were constructed from the photometric fluxes of the AKARI, WISE and IRAS bands, and $2.5-5\,\micron$ spectrum binned with four spectral elements after excluding the spectral data points at the wavelengths of the $\rm H_2O$ ice absorption and Br$\alpha$ emission, because these are not modeled in the {\tt DustEM} code. 
The {\tt DustEM} wrapper outputs the amplitudes of the six components and the abundances of LamC and aSil (hereafter, big grain:~BG), SamC (hereafter, very small grain:~VSG) and PAHs. From the outputs, we calculated $L_{\rm PAH}$ and $L_{\rm TIR}$ which is the total luminosity of BG, VSG and PAHs.

Out of the 57 star-forming galaxies, the observed SEDs of two galaxies in the SDSS sample were not well fitted by model SEDs on a $\chi^2$ test with 90\% confidence and they were excluded from the sample in the following analysis. The resultant sample includes 55 galaxies at spectroscopic redshifts $0.003<z<0.17$ (median redshift $z_{\rm median}\sim0.03$). These are 13 ULIRGs (UltraLuminous InfraRed Galaxies; $L_{\rm TIR}>10^{12}L_{\rm sun}$), 36 LIRGs (Luminous Infrared Galaxies; $10^{11}L_{\rm sun}<L_{\rm TIR}<10^{12}L_{\rm sun}$) and 6 IRGs (InfraRed Galaxies; $L_{\rm TIR}<10^{11}L_{\rm sun}$). All the galaxies in the HST sample are ULIRGs. Figure \ref{fig:sedfit} shows six examples of the result of the SED fitting. 

\subsection{Classification into Merger Galaxies and Non-merger Galaxies}
\begin{figure}
\begin{center}
\includegraphics[width=80mm]{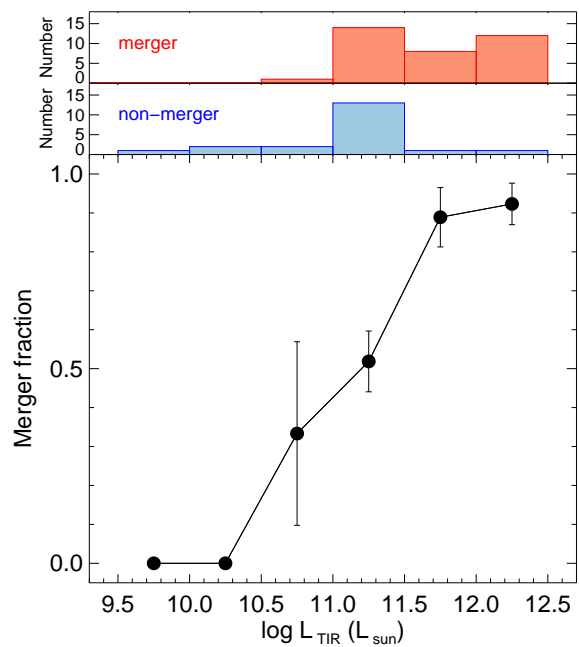}
\caption{
Number fraction of the merger galaxies as a function of $L_{\rm TIR}$. The error bars are based on the binomial statistics. The histograms display the $L_{\rm TIR}$ distributions of the merger and non-merger galaxies. 
}
\label{fig:lir_merger_frac}
\end{center}
\end{figure}

\begin{figure}
\begin{center}
\includegraphics[width=65mm]{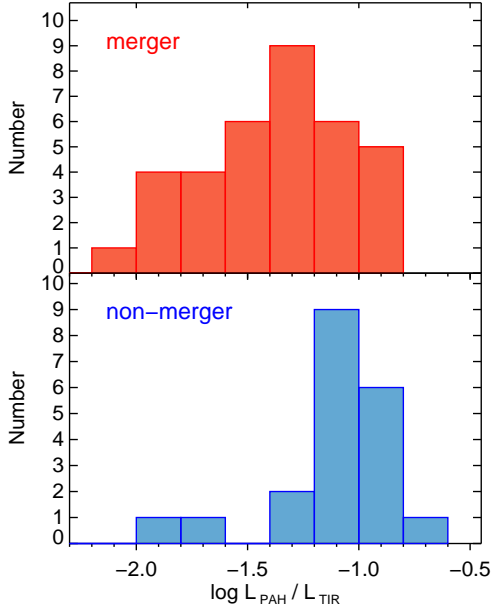}
\caption{
$L_{\rm PAH}/L_{\rm TIR}$ distributions of the merger galaxies (top panel) and the non-merger galaxies (bottom panel). 
}
\label{fig:qpah_tir_dist}
\end{center}
\end{figure}

We divided our sample into merger galaxies and non-merger galaxies using a non-parametric morphological indicator, asymmetry $A$, which is an indicator of morphology asymmetry with $180^\circ$ rotation with respect to the center of a galaxy and has been used in many previous studies to identify merger galaxies (e.g., \citealt{Schade:1995ie}; \citealt{Abraham:1996jm}; \citealt{Lotz:2004gj}; \citealt{Conselice:2008de}; \citealt{Law:2012hf}). We used the SDSS $g$-band images for the 46 galaxies at $z<0.1$ and the HST F814W-band images for the 9 galaxies at $0.1<z<0.2$.

Before calculating $A$, we first created segmentation maps of galaxies as follows: 
\begin{enumerate}
  \item Mask out the pixels of stars and unrelated galaxies in the images. 
  \item Create initial galaxy segmentation maps whose pixels are above $1.5\,\sigma$ sky level and they are connected with the center of the galaxy position. The sky level was adopted as median value of the image after iteratively clipping outliers with the IDL procedure of {\tt mmm.pro}. 
  \item Create the final segmentation maps following the quasi-petrosian image thresholding technique \citep{Abraham:2007hy} which is based on surface brightness distributions in the initial segmentation map with a petrosian fraction of $\eta=0.2$. 
\end{enumerate}
For the 21 galaxies which have a companion galaxy, we define the segmentation map as that including both galaxies. 
The quasi-petrosian image thresholding technique is more suitable for morphological measurement of patchy  galaxies including merger galaxies than the segmentation maps created by the petrosian radius technique. The quasi-petrosian image thresholding technique has been used for many previous studies, especially those at high redshift (e.g., \citealt{Capak:2007aa}; \citealt{Tasca2009aa}; \citealt{Law:2012hf}). 

Then, we calculated $A$. The asymmetry $A$ is defined by
\begin{equation}
  A = \frac{\sum |f_i^{0} - f_i^{180}|}{\sum|f_i^{0}|} - B,
\end{equation}
where $f_i^0$ and $f_i^{180}$ are the fluxes of the $i$-th pixel of the original image and the image rotated by $180^\circ$ about the center, respectively, and $B$ is the correction of the measured asymmetry values due to background fluctuation\footnote{As stated in \citet{Law:2012hf}, there are two commonly used asymmetry definitions which differ by a factor of two. We used the same definition as \citet{Lotz:2004gj}, \citet{Conselice:2008de} and \citet{Law:2012hf}.}. The background correction $B$ is calculated according to the convolution technique by \citet{Zamojski:2007gn}. The summation in the asymmetry $A$ formula is done within the galaxy segment defined above. The rotation center is determined by minimizing asymmetry $A$ around the flux-weighted center in the galaxy segment. 

Figure \ref{fig:thumb-sdss} shows the SDSS $g$-band images of the 46 SDSS galaxies ordered by descending asymmetry from the top left to the bottom right. Figure \ref{fig:thumb-hst} shows the HST F814W images of the 9 HST galaxies. As can be seen in the images, galaxies with large asymmetry values show clear merger signatures such as double nuclei, tidal tails, and/or asymmetric morphology. 

We defined galaxies with $A > 0.35$ as merger galaxies and the other galaxies as non-merger galaxies.
The adopted asymmetry criterion of 0.35 is the same as in \citet{Conselice:2003ds}. As a result, out of the SDSS 46 galaxies, 26 galaxies ($\sim57\%$) are classified as merger galaxies and the other 20 galaxies as non-merger galaxies ($\sim43\%$), while all the 9 HST galaxies are classified as merger galaxies. Out of all the 55 galaxies, 35 galaxies ($\sim64\%$) are classified as merger galaxies and the other 20 galaxies as non-merger galaxies ($\sim36\%$). 
The merger galaxies consist of galaxies with a companion and those with evidence of recent mergers such as double nuclei, tidal tails and/or asymmetric morphology in a single envelope. 

\begin{figure}
\begin{center}
\includegraphics[width=65mm,angle=-90]{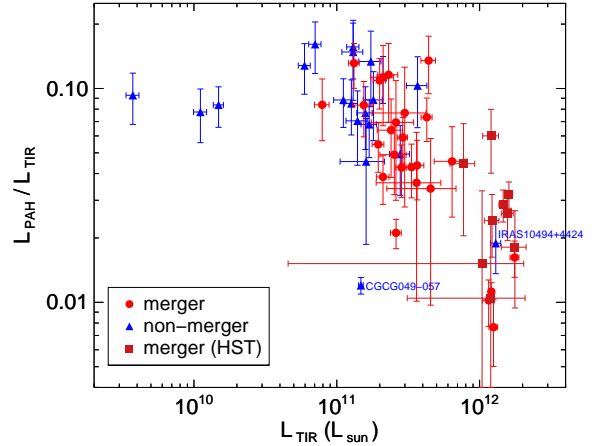}
\caption{
$L_{\rm PAH}/L_{\rm TIR}$ as a function of $L_{\rm TIR}$ for the merger galaxies (red circles) and the non-merger galaxies (blue triangles) in the SDSS sample, and merger galaxies (red squares) in the HST sample. 
}
\label{fig:lir_qpah}
\end{center}
\end{figure}

\section{Results}
\subsection{Relation to galaxy merger} 

\begin{figure*}
\begin{center}
\includegraphics[width=55mm,angle=-90]{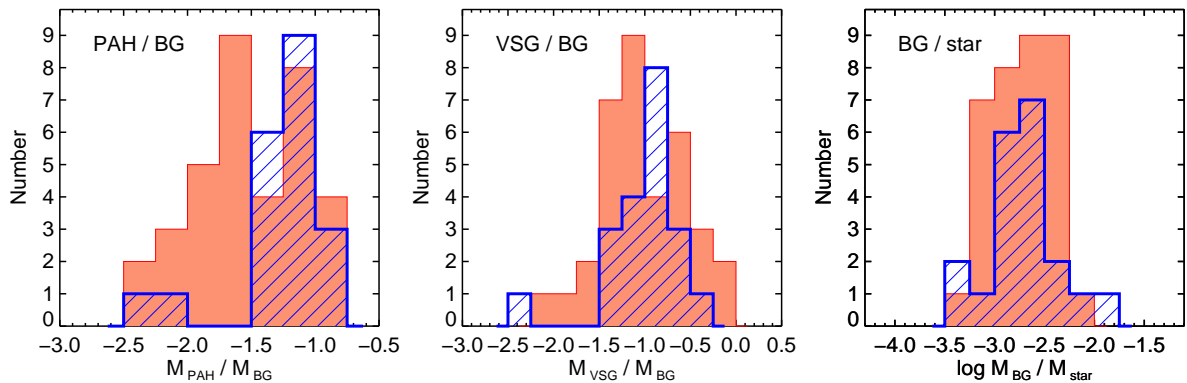}
\caption{
Distributions of $M_{\rm PAH}/M_{\rm BG}$, $M_{\rm VSG}/M_{\rm BG}$ and $M_{\rm BG}/M_{\rm star}$ for merger galaxies (filled red histogram) and non-merger galaxies (shaded blue histogram). 
}
\label{fig:hist_mq3}
\end{center}
\end{figure*}

\begin{figure}
\begin{center}
\includegraphics[width=85mm,angle=-90]{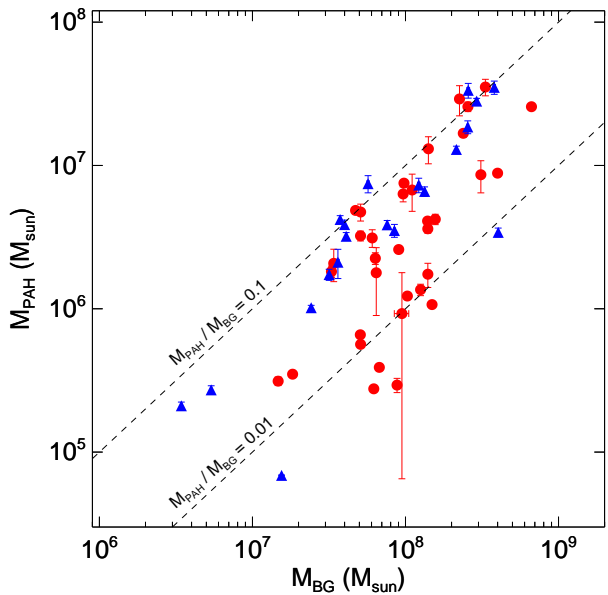}
\caption{
$M_{\rm PAH}$ as a function of $M_{\rm BG}$ for merger galaxies (red circles) and non-merger galaxies (blue triangles). The error bars are generated by the SED fitting of the {\tt DustEM} with the observed flux errors.  
} 
\label{fig:mass}
\end{center}
\end{figure}

Figure \ref{fig:lir_merger_frac} displays a number fraction of the merger galaxies as a function of $L_{\rm TIR}$, which shows that the fraction increases with $L_{\rm TIR}$. The merger galaxies occupy $\sim17\%$ for IRGs, $\sim61\%$ for LIRGs, and $\sim92\%$ for ULIRGs, suggesting that $L_{\rm TIR}$ of most ULIRGs and almost half LIRGs are mainly induced by galaxy mergers. Since our sample is pure star-forming galaxies with no AGN signatures, $L_{\rm TIR}$ of the merger galaxies should be mainly caused by merger-induced starburst. The fraction is broadly consistent with the previous studies of local galaxies (e.g., \citealt{Sanders:1996cd}; \citealt{Kim:2002hc}; \citealt{Veilleux:2002dj}; \citealt{Wang:2006hu}; \citealt{KilerciEser:2014wv}). The fraction of interacting/merger systems increases from $\sim10\%$ at $L_{\rm TIR}=10^{10.5}-10^{11}~L_{\rm sun}$ to $\sim100\%$ at $L_{\rm TIR}>10^{12}~L_{\rm sun}$ by visual inspection of the IRAS Bright Galaxy Survey (\citealt{Sanders:1996cd}). \citet{Wang:2006hu} found that $48\%$ of LIRGs at $z\sim0.05$ are interacting/merging by visual inspection of the SDSS $r$-band and composite color images. All the ULIRGs but one in the IRAS 1 Jy sample show signs of a strong tidal interaction and merger by visual inspection of the optical images (\citealt{Kim:2002hc}; \citealt{Veilleux:2002dj}). \citet{KilerciEser:2014wv} recently found that ULIRGs in the AKARI all-sky survey are interacting pair galaxies or on-going/post mergers by visual inspection of the SDSS color composite images. 

Figure \ref{fig:qpah_tir_dist} displays that $L_{\rm PAH}/L_{\rm TIR}$ of the merger galaxies tends to be lower than that of the non-merger galaxies. 
The median $L_{\rm PAH}/L_{\rm TIR}$ of the merger galaxies and the non-merger galaxies are 0.044 and 0.088, respectively. 
Two non-merger galaxies of CGCG\,049-057 and IRAS\,10494+4424 have extremely low $L_{\rm PAH}/L_{\rm TIR}$. \citet{Peeters:2004ga} also reported that CGCG\,049-057 has a low luminosity ratio of the PAH $6.2\,\micron$ emission to FIR continuum due to its unusual infrared spectrum with strong PAH $6.2\,\micron$ emission, and very strong and cold FIR continuum. 
IRAS\,10494+4424 is close to our asymmetry criterion of the classification between merger galaxies and non-merger galaxies. Asymmetric diffuse light distribution, which implies that this galaxy undergoes a recent merging process, is recognized in the $g$-band image of IRAS\,10494+4424 by visual inspection, and the galactic nucleus is extremely brighter than the diffuse light. Thus this galaxy may be mistakenly classified into non-merger galaxies. 
We performed the Mann-Whitney $U$ test and rejected with a significant level of $5\%$ the null hypothesis that the $L_{\rm PAH}/L_{\rm TIR}$ distributions of the merger and the non-merger galaxies are drawn from the distribution with the same median. We therefore confirm that $L_{\rm PAH}/L_{\rm TIR}$ of the merger galaxies is statistically lower than that of the non-merger galaxies. 

Figure \ref{fig:lir_qpah} shows the relationship between $L_{\rm PAH}/L_{\rm TIR}$ and $L_{\rm TIR}$ for merger galaxies and non-merger galaxies. The merger galaxies and the non-merger galaxies for the SDSS sample are plotted as red circles and blue triangles respectively, while the merger galaxies for the HST sample are plotted as red squares. Note that there are no non-merger galaxies in the HST sample, which reflects that all the galaxies in the HST sample are ULIRGs. 
Above $L_{\rm TIR}>10^{11} L_{\rm sun}$, $L_{\rm PAH}/L_{\rm TIR}$ clearly decreases with $L_{\rm TIR}$ from $\sim0.10$ at $L_{\rm TIR}\sim10^{11} L_{\rm sun}$ to $\sim0.01$ at $L_{\rm TIR}\sim10^{12} L_{\rm sun}$, which is consistent with the previous studies that the ratio of the PAH feature luminosity to $L_{\rm TIR}$ decreases with increasing $L_{\rm TIR}$ in the local universe and at high redshift (\citealt{Elbaz:2011ix}; \citealt{Nordon:2012io}; \citealt{Yamada:2013ca}; \citealt{Murata:2014hy}; \citealt{Stierwalt:2014aa}). 
In Figure \ref{fig:lir_qpah}, we find that the decrease is mainly due to lower $L_{\rm PAH}/L_{\rm TIR}$ of merger galaxies than non-merger galaxies. 
A similar trend was reported in \citet{Stierwalt:2014aa} who investigated $L_{\rm PAH}/L_{\rm TIR}$ with the stages of merging processes for U/LIRGs using 5-38 $\micron$ spectra taken with {\it Spitzer} Infrared Spectrograph (IRS) in the Great Observatories All-sky LIRG Survey (GOALS). \citet{Stierwalt:2014aa} classified each galaxy into five classes (non-mergers, pre-mergers, early-stage mergers, mid-stage mergers, late-stage mergers) with visual inspection of the {\it Spitzer} IRAC 3.6\,$\micron$ images performed in \citet{Stierwalt:2013aa} and found the decrease in $L_{\rm PAH}/L_{\rm TIR}$ for late-stage mergers compared to pre-mergers, early-stage mergers, non-mergers. Considering that late-stage mergers tend to have lower equivalent width of 6.2\,$\micron$ PAH emission than pre-mergers, the decrease is considered as an excess of $L_{\rm TIR}$ not associated with star formation, namely warm dust emission of AGNs.  
In our study, we find that merger galaxies have lower $L_{\rm PAH}/L_{\rm TIR}$ than non-merger galaxies even for pure star-forming galaxies, which suggests that AGN emission may not contribute to the decrease. 

Figure \ref{fig:hist_mq3} shows the ratios of $M_{\rm PAH}/M_{\rm BG}$, $M_{\rm VSG}/M_{\rm BG}$ and $M_{\rm BG}/M_{\rm star}$ for merger galaxies and non-merger galaxies. 
Here BG mass $\left(M_{\rm BG}\right)$, VSG mass $\left(M_{\rm VSG}\right)$ and PAH mass $\left(M_{\rm PAH}\right)$ are calculated from the results of the SED fitting with the {\tt DustEM}. Stellar mass $\left(M_{\rm star}\right)$ is estimated from the WISE $3.4\,\micron$ and $4.6\,\micron$ band flux densities using the conversion from WISE NIR photometry and a redshift to a stellar mass given by \citet{Eskew:2012ft}. We adopted the Chabrier initial mass function \citep{Chabrier:2003aa}. The W1-band and W2-band fluxes are adopted as aperture photometry with the aperture size of $24\farcs75$ in the radius in the WISE all sky source catalog. The error bars of $M_{\rm PAH}$, $M_{\rm VSG}$, $M_{\rm BG}$ are generated by the SED fitting of the {\tt DustEM} with the observed flux errors. The error bars of $M_{\rm star}$ are estimated by the propagation of the errors of the WISE flux densities. 

Figure \ref{fig:hist_mq3} clearly shows that merger galaxies tend to have lower $M_{\rm PAH}/M_{\rm BG}$ than non-merger galaxies. Almost half of the merger galaxies have lower $M_{\rm PAH}/M_{\rm BG}$ ratios than non-merger galaxies although the other half of the merger galaxies have $M_{\rm PAH}/M_{\rm BG}$ similar to non-merger galaxies. By contrast, there are no systematic differences in $M_{\rm VSG}/M_{\rm BG}$ between merger galaxies and non-merger galaxies. Also, merger galaxies have $M_{\rm BG}/M_{\rm star}$ similar to non-merger galaxies. 
The systematic decrease of $M_{\rm PAH}/M_{\rm BG}$ for merger galaxies are clearly seen in Figure \ref{fig:mass} which displays $M_{\rm PAH}$ as a function of $M_{\rm BG}$. Merger galaxies clearly have lower $M_{\rm PAH}$ at a given $M_{\rm BG}$ than non-merger galaxies. These results indicate that the tendency of low $M_{\rm PAH}/M_{\rm BG}$ of merger galaxies is caused by low $M_{\rm PAH}$ of merger galaxies, which suggests a part of PAH destruction during merging processes of galaxies. 

\subsection{Relation to the stages of merging processes} 
The merger galaxies do not always possess lower $M_{\rm PAH}/M_{\rm BG}$ than non-merger galaxies. 
Obviously, the next question is what causes the difference in $M_{\rm PAH}/M_{\rm BG}$ for merger galaxies. By our definition on merger galaxies which are classified with $A$, the merger galaxies can include a wide range of merging stages. Therefore, it is interesting to examine the stages of merging processes for the merger galaxies and investigate the relation of PAH abundance with the stages of merging processes. 

Motivated by this, we estimate two morphological indicators, the Gini coefficient $G$ and $M_{20}$ \citep{Lotz:2004gj}, and investigate their relation to $M_{\rm PAH}/M_{\rm BG}$. $G$ means inequality of a flux distribution, and $M_{20}$ is a logarithmic fraction of the intensity momentum of the brightest region which includes 20\% flux of the total flux to the momentum of the total region of a galaxy. The definitions of $G$ and $M_{20}$ are described in the Appendix. Since galaxies at different merging stages occupy different regions of $G$ and $M_{20}$ diagram \citep{Lotz:2008wg}, we can investigate stages of merger galaxies. 

\begin{figure}
\begin{center}
\includegraphics[width=70mm,angle=-90]{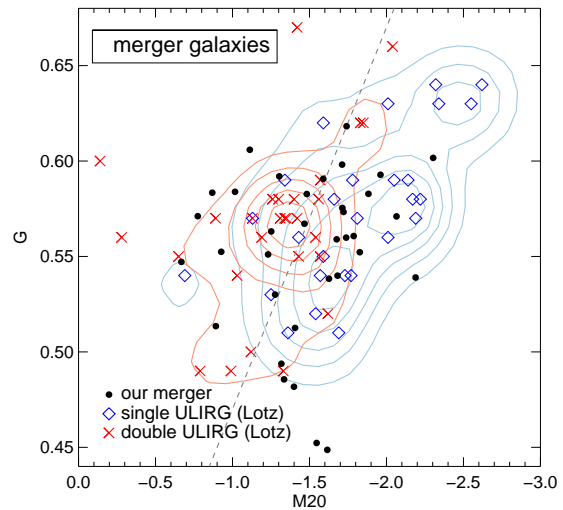}
\caption{
$G$ and $M_{20}$ distribution for merger galaxies. The filled circles indicate our sample of merger galaxies. For reference, we plot a $G$-$M_{20}$ distribution of ULIRGs taken from \citet{Lotz:2004gj} with the conversion from $R$ band to $g$ band. The diamonds and the blue contour indicate $G$ and $M_{20}$ of the single ULIRGs and their number density, respectively. The crosses and the red contour indicate $G$ and $M_{20}$ of double ULIRGs and their number density, respectively. The contours are drawn with linearly spaced five levels. The dashed line indicates the best classification line for two classes of the single and double ULIRGs. 
}
\label{fig:gm20_A}
\end{center}
\end{figure}

Figure \ref{fig:gm20_A} shows a distribution of $G$ and $M_{20}$ for our merger galaxies. For reference, we plot $G$-$M_{20}$ distributions of ULIRGs with single nucleus (single ULIRGs) and ULIRGs with double nuclei (double ULIRGs) taken from \citet{Lotz:2004gj}. Since the $G$ and $M_{20}$ of the ULIRGs of \citet{Lotz:2004gj} were calculated in $R$ band images, we added $M_{20}$ value by $0.1$ and $G$ by $-0.02$, which are typical differences between the two bands taken from \citet{Lotz:2004gj}, to convert from $R$ band to $g$ band used in the $G$ and $M_{20}$ calculation of our merger galaxies. We performed Fisher's linear discriminant analysis for two classes of the single and double ULIRGs, and plot the best classification line ($G=-0.20M_{20}+0.27$) in Figure \ref{fig:gm20_A}. 
Since ULIRGs in the local universe are considered to be merger galaxies of gas-rich disk galaxies on the basis of their morphology and starburst activities \citep{Veilleux:2002dj}, their $G$-$M_{20}$ distribution is used as a representative of merger galaxies \citep{Lotz:2004gj}. Based on numerical simulations of galaxy mergers, it is considered that single ULIRGs are at late stages of merging processes of galaxies including the final coalescence stage and double ULIRGs are earlier stages than single ULIRGs (\citealt{Veilleux:2002dj}). 
Therefore, merger galaxies can be divided into two classes using the best classification line in the $G-M_{20}$ distribution of Figure \ref{fig:gm20_A}. Merger galaxies at early (late) stages are those above (below) the line. 

Although our merger galaxies are, for the first approximation, similarly distributed in the $G$-$M_{20}$ region of the ULIRGs, there shows a little discrepancy at $G\sim0.65$ and $M_{20}\sim-2.5$; no galaxies are found in our sample while the local number density peak is seen for the single ULIRGs. The discrepancy may be because our sample of merger galaxies tends to possess lower $L_{\rm TIR}$ than the ULRGs of \citet{Lotz:2004gj}. Merger-induced starburst is considered to occur at nuclear region. Galaxies with higher $L_{\rm TIR}$, that is higher SFR, may possess more luminous nuclear starburst, which may cause higher $G$ and lower $M_{20}$. 
\citet{Psychogyios:2016aa} examined a distribution of $G$ and $M_{20}$ for 89 LIRGs from the GOALS sample. Their $M_{20}$ is similar to our result. On the other hand, their $G$ is much smaller than ours and \citet{Lotz:2004gj}, which may be due to their small image size as pointed out by them. 

\begin{figure}
\begin{center}
\includegraphics[width=70mm,angle=-90]{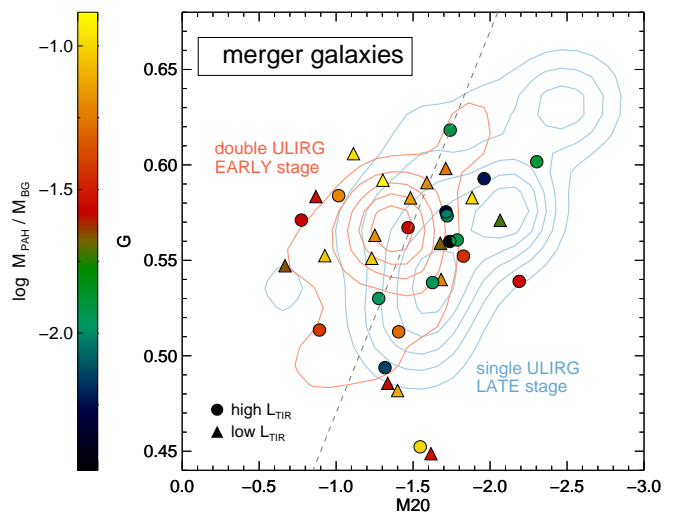}
\caption{
$M_{\rm PAH}/M_{\rm BG}$ variation of merger galaxies in $G$ and $M_{20}$ diagram. The colors scale with $\log M_{\rm PAH}/M_{\rm BG}$. The contours and the dashed line are the same as in Figure \ref{fig:gm20_A}. The sample is divided into two equal sized bins with their $L_{\rm TIR}$: (circle) galaxies with $L_{\rm TIR} > 3.63\times10^{11}L_{\rm sun}$, (triangle) galaxies with $L_{\rm TIR} <3.63\times10^{11}L_{\rm sun}$. 
}
\label{fig:gm20_lir}
\end{center}
\end{figure}

Figure \ref{fig:gm20_lir} displays a variation of $M_{\rm PAH}/M_{\rm BG}$ of our merger galaxies in $G$-$M_{20}$ diagram and shows a trend of the PAH destruction with the stages of merging processes. 
We find that the merger galaxies with relatively low $M_{20}$ tend to have low $M_{\rm PAH}/M_{\rm BG}$ than those with relatively high $M_{20}$. Galaxies with $M_{\rm PAH}/M_{\rm BG}<-1.7$, where non-merger galaxies are few, are clustered around the region occupied by the single ULIRGs. Out of the 11 galaxies with such very low $M_{\rm PAH}/M_{\rm BG}$, nine galaxies have lower $M_{20}$ against the classification line and even the rest two galaxies have a little higher $M_{20}$ against the line. 
Furthermore, we divided the sample into two equal sized bins with their $L_{\rm TIR}$ and found that, for the galaxies with high $L_{\rm TIR}$, galaxies with very low $M_{\rm PAH}/M_{\rm BG}$ are also seen in the region occupied by the single ULIRGs. 
The trend indicates that the galaxies with such very low $M_{\rm PAH}/M_{\rm BG}$ are merger galaxies at late stages, reflecting the tendency of lower $M_{\rm PAH}/M_{\rm BG}$ for merger galaxies as seen in Figure \ref{fig:hist_mq3}.

\begin{figure}
\begin{center}
\includegraphics[width=70mm,angle=0]{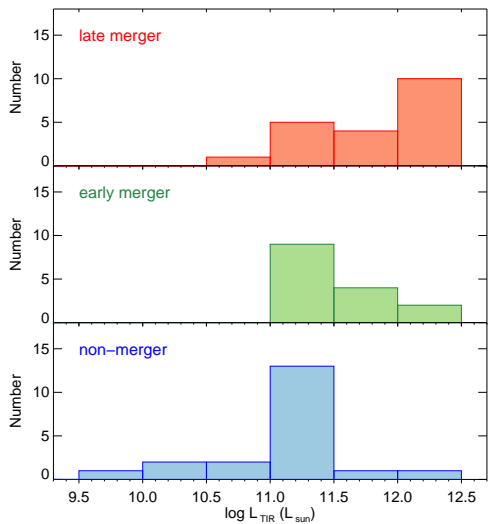}
\caption{
$L_{\rm TIR}$ distributions of the merger at late stages (top), those at early stages (middle) and non-merger galaxies (bottom). 
}
\label{fig:lir_merger_stage}
\end{center}
\end{figure}

Figure \ref{fig:lir_merger_stage} shows $L_{\rm TIR}$ distributions of the merger galaxies at late stages, those at early stages and non-merger galaxies. Hereafter, merger galaxies at early (late) stages are defined as those above (below) the best classification line in the $G-M_{20}$ distribution of Figure \ref{fig:gm20_A}. The stages of merging processes are clearly related with $L_{\rm TIR}$. Merger galaxies at late stages possess higher $L_{\rm TIR}$ than those at early stages. Ten out of 13 ULIRGs ($\sim77\%$) are merger galaxies at late stages. More than half of merger galaxies with $10^{11.0}<L_{\rm TIR}<10^{11.5}$ are at early stages. These results are broadly consistent with the previous study of \citet{Larson:2016aa} despite the different method of merger classification. 
\citet{Larson:2016aa} performed visual inspection of the {\it HST} $I$-band images for 65 local luminous infrared galaxies of the GOALS sample and investigated the relation of merger stages and $L_{\rm TIR}$. They found that almost all ULIRGs are major merger galaxies at late stages while more than half of merger galaxies with $10^{11.1}<L_{\rm TIR}<10^{11.5}$ are at early stages. 

% %------------------------------------------------------------------------------
% %    Discussion
% %------------------------------------------------------------------------------
\section{Discussion} 
Using the AKARI, WISE, IRAS, SDSS and HST data, we investigated the relation of $L_{\rm TIR}$ and $L_{\rm PAH}$ with galaxy merger for 55 star-forming galaxies taken from the sample of \citet{Yamada:2013ca}. 
As a result, we confirm the previously noted trend that the fraction of merger galaxies increases with $L_{\rm TIR}$ and merger galaxies tend to have lower $L_{\rm PAH}/L_{\rm TIR}$ than non-merger galaxies, which results in the decreases of $L_{\rm PAH}/L_{\rm TIR}$ with increasing $L_{\rm TIR}$ (Figures \ref{fig:lir_merger_frac}, \ref{fig:qpah_tir_dist} and \ref{fig:lir_qpah}).  

Furthermore, we investigated the relation of $M_{\rm PAH}$, $M_{\rm VSG}$, $M_{\rm BG}$ and $M_{\rm star}$ with galaxy merger (Figures \ref{fig:hist_mq3} and \ref{fig:mass}). We find that merger galaxies tend to have lower $M_{\rm PAH}/M_{\rm BG}$ than non-merger galaxies. Almost a half of the merger galaxies have lower $M_{\rm PAH}/M_{\rm BG}$ ratios than non-merger galaxies, while the other half of the merger galaxies have $M_{\rm PAH}/M_{\rm BG}$ similar to non-merger galaxies. By contrast, there are no systematic differences in $M_{\rm VSG}/M_{\rm BG}$ between merger galaxies and non-merger galaxies. Also, merger galaxies have $M_{\rm BG}/M_{\rm star}$ similar to non-merger galaxies. 
These results suggest that PAHs are partly destroyed during merging processes of galaxies. 
Figure \ref{fig:gm20_lir} suggests that PAHs are destroyed especially at the late stages of merging processes. 

The destruction of PAHs in U/LIRGs has been reported by previous studies and it is suggested to be caused by harsh radiation fields of AGNs (e.g., \citealp{Stierwalt:2013aa,Stierwalt:2014aa}). Since galaxies contaminated by AGN activities were strictly excluded from our sample by two methods of the previously established NIR AGN diagnostics as explained in Section 2, the PAH destruction found in our sample needs another destruction processes.

We consider possible origins for PAH destruction during merging processes of galaxies. The first possibility is the photo-dissociation of PAHs by strong FUV radiation field due to star-formation activity (e.g., \citealt{Murata:2014hy}, \citealt{Domingue:2016aa}). PAHs can be destroyed by absorption of some Far-UV (FUV: $6-13.6$~eV) photons since such events heat PAHs above their sublimation temperature. It is well known that merging processes of gas-rich galaxies can trigger starburst (e.g., \citealt{Mihos:1996aa}; \citealt{Springel:2005aa}; \citealt{Cox:2008aa}), and thus strong FUV radiation field, which may cause the decrease of $M_{\rm PAH}/M_{\rm BG}$ for merger galaxies relative to non-merger galaxies. 

Second, possible origin of the PAH destruction is large-scale shocks by merging processes of galaxies. 
\citet{Yamada:2013ca} suggested that low abundance ratios of PAHs to dust are due to PAH destruction by large-scale shocks during merging processes of galaxies. Merging of galaxies is considered to induce large-scale strong shocks in the galaxies (e.g., \citealp{Rich:2015kf}) and these shocks can destroy PAHs. The destructions of PAHs under the shocks depend on the PAH size and the shock speed. Although shocks with the speed $v_{\rm shock}<150$~km/s destroy only small PAHs, shocks with $v_{\rm shock}>150$~km/s destroy almost all PAHs, regardless of their sizes \citep{Micelotta:2010dl}. On the other hand, shocks with $v_{\rm shock}\sim200$~km/s destroy dust grains with only $10-20\,\%$ dust masses \citep{Jones:1996bi}. LIRGs and ULIRGs possess extended ionized gas with high velocities dispersions, which are interpreted as large-scale shocks with $v_{\rm shock}\sim150-500$~km/s induced by galaxy mergers \citep{MonrealIbero:2006fi,MonrealIbero:2010gz,Rich:2011is,Rich:2014ib,Rich:2015kf}. 
Recently, \citet{Rich:2014ib,Rich:2015kf} performed integral field spectroscopic (IFS) observations for 27 nearby U/LIRGs of the GOALS sample and found that the fraction of H$\alpha$ emission with high velocity dispersion $>90$~km/s, which is mainly originated by shock excitation, to the total H$\alpha$ emission for merger galaxies statistically increases with the stages of merging processes. The fraction increases from about $10\,\%$ at pair galaxies with wide separations of $10-100$\,kpc to about $35\,\%$ at close pair galaxies with separations of $<10$\,kpc, and finally reaches almost $60\,\%$ at the final coalescence stage. 

Post-shock ionized gas now emitting the optical emission lines is possible to be pre-shock dense gas such as that in photodissociation regions (PDRs) and molecular clouds where PAHs are present since the shocks can dissociate molecules and ionize atomic gas. In this case, the presence of the shocked gas detected by optical emission lines means that PAHs can be under shocks during galaxy mergers and thus they can be partly destroyed. Furthermore, since galaxy mergers can induce large-scale shocks by tidally-induced high velocity flows and cloud-cloud collision of dense gas cloud, the PDRs in merger galaxies may be also under the shocks. 
Since the sample of \citet{Rich:2014ib,Rich:2015kf} consists of typical local U/LIRGs, large-scale shocks commonly appear in local U/LIRGs with merging processes and possibly in gas-rich major mergers. Although local galaxies with large-scale shocks are extremely rare for the overall galaxy population due to the rarity of gas-rich major merger galaxies, our sample of U/LIRGs mostly during merging processes is expected to show large-scale shocks. Under the PAH destruction scenario by large-scale shocks during merging processes of galaxies, this result implies that PAHs are expected to be destroyed more efficiently at later stages, especially the final coalescence stage, during merging processes. 

Finally, it is possible that both strong UV radiation and large-scale shocks comparably contribute to the PAH destruction since merging processes of galaxies can induce both. 

% ----------------------------------------------------------
%                  verification
% ----------------------------------------------------------
To verify the above possibilities, we investigate $M_{\rm PAH}/M_{\rm BG}$ variations in $G_0$ and the emission line ratio of $[\ion{O}{i}]\lambda6300/{\rm H}\alpha$ (hereafter denoted $[\ion{O}{i}]/{\rm H}\alpha$) for the merger galaxies. 
Here $G_0$ is radiation field intensity integrated for the FUV wavelengths relative to the standard InterStellar Radiation Field (ISRF) and is derived from the results of the SED fitting with the {\tt DustEM}. 
Optical emission line ratios such as $[\ion{O}{i}]/{\rm H}\alpha$ are used to distinguish the shocks from \ion{H}{ii} regions (e.g., \citealt{Lequeux:2005aa}). 
The large $[\ion{O}{i}]/{\rm H}\alpha$ is considered as a evidence of galaxy-wide shocks for several optical IFS observations (\citealt{Farage:2010aa}; \citealt{MonrealIbero:2010gz}; \citealt{Rich:2011is}; \citealt{Rich:2014ib}; \citealt{Rich:2015kf}). 
Since the emission line ratio of $[\ion{O}{i}]/{\rm H}\alpha$ is more pronounced shock tracer than those of [\ion{S}{ii}]$\lambda6717,6731$/H$\alpha$ and [\ion{N}{ii}]$\lambda 6584$/H$\alpha$ \citep{MonrealIbero:2010gz}, we determined to here utilize the emission line ratio of $[\ion{O}{i}]/{\rm H}\alpha$ for a shock tracer. 
Although the spatial distribution of $[\ion{O}{i}]/{\rm H}\alpha$ is strong evidence of shocks, it is not available in our sample. Therefore, the emission line of spatially-integrated spectroscopic data is used in the following analysis. 

We retrieved the [\ion{O}{i}] and H$\alpha$ line fluxes of the merger galaxies in our sample from archival data of spectroscopic observations and constructed the emission line ratios of $[\ion{O}{i}]/{\rm H}\alpha$. For the galaxies with a companion galaxy, we used the emission line ratios of primary galaxies because few companion galaxies have spectroscopic data. 
Out of the 35 merger galaxies in our sample, the emission line ratios of $[\ion{O}{i}]/{\rm H}\alpha$ of 23 galaxies are constructed using the SDSS fiber spectroscopic data (MPA-JHU catalog, ver. 5.2 based on the SDSS 7th data). Out of the 12 merger galaxies without the SDSS data, 5 galaxies are constructed using the spectroscopic data by \citet{Veilleux:1999aa}. The rest 7 galaxies have no spectroscopic data and they are not included in the following analysis. Seven galaxies out of the 23 galaxies with the SDSS data also have the emission line fluxes of [\ion{O}{i}] and H$\alpha$ of \citet{Veilleux:1999aa}. To assess the systematic offset of $\log$$[\ion{O}{i}]/{\rm H}\alpha$ of the two spectroscopic data, we calculated the mean difference of $\log$$[\ion{O}{i}]/{\rm H}\alpha$ between the SDSS and \citet{Veilleux:1999aa} data. The difference is 0.08 dex and is negligible to our result.  

\begin{figure}
\begin{center}
\includegraphics[width=63mm,angle=-90]{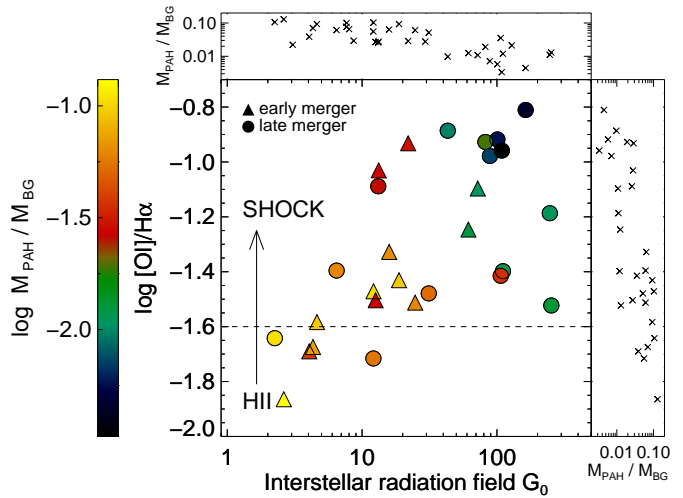}
\caption{
$M_{\rm PAH}/M_{\rm BG}$ variation in the interstellar radiation field $G_0$ and [\ion{O}{i}]$\lambda6300$/H$\alpha$ diagram for the merger galaxe with [\ion{O}{i}]$\lambda6300$ and H$\alpha$ emission lines available. Merger galaxies at early stages and those at late stages are triangles and circles, respectively. 
The scatter plots shown in the main panel indicate $M_{\rm PAH}/M_{\rm BG}$ variation as a function of $G_0$ and [\ion{O}{i}]$\lambda6300$/H$\alpha$. The dashed line is distinguish from the shocks from \ion{H}{ii} regions \citep{MonrealIbero:2010gz}. 
Merger galaxies at early (late) stages are defined as those above (below) the best classification line in the $G-M_{20}$ distribution of Figure \ref{fig:gm20_A}. 
}
\label{fig:G0_OIHa_mqpahbg}
\end{center}
\end{figure}

Figure \ref{fig:G0_OIHa_mqpahbg} shows $M_{\rm PAH}/M_{\rm BG}$ variation in the $G_0$ and $[\ion{O}{i}]/{\rm H}\alpha$ emission line ratio diagram for the 28 merger galaxies with [\ion{O}{i}] and H$\alpha$ emission line fluxes available. 
Galaxies with $\log[\ion{O}{i}]/{\rm H}\alpha>-1.6$ are best explained by the presence of shocks with $v_{\rm shock}<200$~km/s, while $\log[\ion{O}{i}]/{\rm H}\alpha<-1.6$ is explained by the emission from gas ionized by young stars in \ion{H}{ii} region \citep{MonrealIbero:2010gz}. Most of the galaxies occupy the shocked region. Merger galaxies at late stages tend to be larger $[\ion{O}{i}]/{\rm H}\alpha$ than those at early stages, which is consistent with the IFU surveys that shocks are an increasingly important component of the optical emission lines as merging processes \citep{MonrealIbero:2010gz,Rich:2014ib,Rich:2015kf}. 

From Figure \ref{fig:G0_OIHa_mqpahbg}, it is clear that $M_{\rm PAH}/M_{\rm BG}$ decreases with increasing $G_0$ and $[\ion{O}{i}]/{\rm H}\alpha$. Interestingly, galaxies at a given $G_0$ tend to possess lower $M_{\rm PAH}/M_{\rm BG}$ with increasing $[\ion{O}{i}]/{\rm H}\alpha$, which suggests that PAHs tend to be destroyed more in shock-dominated galaxies even with the same  radiation field strength. Furthermore, galaxies at a given $[\ion{O}{i}]/{\rm H}\alpha$ tend to possess lower $M_{\rm PAH}/M_{\rm BG}$ with increasing $G_0$, which suggests that radiation fields strength is another important factor for the PAH destruction. Therefore, it is likely that both large-scale shocks and strong radiation fields comparably contribute to destroy PAHs during merging processes of galaxies. 

\begin{figure}
\begin{center}
\includegraphics[width=85mm,angle=0]{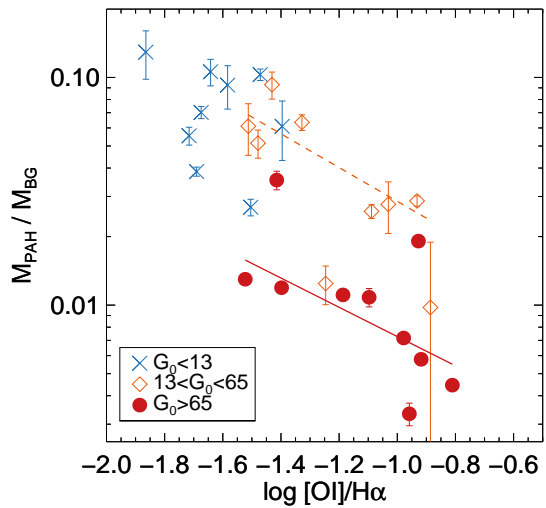}
\caption{
$M_{\rm PAH}/M_{\rm BG}$ as a function of [\ion{O}{i}]$\lambda6300$/H$\alpha$ for the same sample of Figure \ref{fig:G0_OIHa_mqpahbg}. The sample is divided into three equally-sized bins with their $G_0$: galaxies with  (cross) $G_0<13$, (diamond) $13<G_0<65$ and (circle) $G_0>65$. The lines represent the fitting results of galaxies with $13<G_0<65$ (dashed) and $G_0>65$ (solid) with a linear line of $\log(M_{\rm PAH}/M_{\rm BG})=a\times\log([\ion{O}{i}]\lambda6300/{\rm H}\alpha)+b$, where $a$ and $b$ are free parameters. 
}
\label{fig:OIHa_mqpahbg}
\end{center}
\end{figure}

These trends are clearly seen in Figure \ref{fig:OIHa_mqpahbg}, which shows $M_{\rm PAH}/M_{\rm BG}$ as a function of the $[\ion{O}{i}]/{\rm H}\alpha$ line ratio at a given $G_0$ for the same sample of Figure \ref{fig:G0_OIHa_mqpahbg}. The sample is divided into three equal sized bins according to $G_0$; galaxies with low $G_0$: $G_0<13$, intermediate $G_0$: $13<G_0<65$, and high $G_0$: $G_0>65$. 
For the merger galaxies with intermediate and high $G_0$, $M_{\rm PAH}/M_{\rm BG}$ decreases with the $[\ion{O}{i}]/{\rm H}\alpha$ line ratio. The $M_{\rm PAH}/M_{\rm BG}$ of the galaxies with intermediate (high) $G_0$ decreases from $\sim0.06$ ($\sim0.01$) at $\log[\ion{O}{i}]/{\rm H}\alpha\sim-1.5$ to $\sim0.03$ ($\sim0.007$) at $\log[\ion{O}{i}]/{\rm H}\alpha\sim-1.0$. The Spearman's rank test found negative correlations between $M_{\rm PAH}/M_{\rm BG}$ and $[\ion{O}{i}]/{\rm H}\alpha$ for galaxies with intermediate $G_0$ (Spearman's correlation coefficient $\rho=-0.70$) and high $G_0$ ($\rho=-0.66$) with a significant level of $5\%$. On the other hand, no significant correlations for galaxies with low $G_0$ are found with the Spearman's rank test. We fit the data points of galaxies with intermediate and high $G_0$ with a linear line of $\log(M_{\rm PAH}/M_{\rm BG})=a\times\log([\ion{O}{i}]/{\rm H}\alpha)+b$, where $a$ and $b$ are free parameters. 
The best-fit linear lines shown in Figure \ref{fig:G0_OIHa_mqpahbg} have similar slopes each other with a systematic offset in $M_{\rm PAH}/M_{\rm BG}$. Galaxies with high $G_0$ have lower $M_{\rm PAH}/M_{\rm BG}$ than those with intermediate $G_0$ at a given $[\ion{O}{i}]/{\rm H}\alpha$ line ratio. 
In summary, PAH destruction is likely to be caused by two processes; large-scale shocks and strong interstellar radiation fields during merging processes of galaxies especially at late stages. 

For major-merger galaxy pairs at $z\sim0$, \citet{Domingue:2016aa} found that PAH-to-dust mass ratios of spiral-spiral pairs are lower than those of spiral-elliptical pairs and isolated spiral galaxies. They suggested that the reduction is likely a PAH deficiency by their enhanced interstellar radiation fields caused by their active star formation. The sample of \citet{Domingue:2016aa} is mostly IRGs, while our sample of merger galaxies consists of mostly LIRGs and ULIRGs including merger galaxies at later stages. IRGs have lower radiation field strength than U/LIRGs. Main processes of PAH destruction might differ according to $L_{\rm TIR}$ and the stages of merging processes. PAHs might be destroyed by strong radiation fields by star formation in IRGs, while they might be destroyed by both strong radiation fields and large-scale shocks in LIRGs and ULIRGs which are dominant for merger galaxies. 

\vspace{5mm}
Many authors pointed out the deficit of [\ion{C}{ii}] 158$\,\micron$ emission line for star-forming and starburst galaxies including IRGs, LIRGs and ULIRGs, the so-called [\ion{C}{ii}] deficit. (e.g., \citealp{Luhman:2003aa}). Since [\ion{C}{ii}] and PAH emissions are mainly emitted from PDRs and they are tied to each other (\citealt{Diaz-Santos:2013aa}; \citealt{Diaz-Santos:2014aa}), it is worth noting the relation between the PAH destruction during merging processes of galaxies and the [\ion{C}{ii}] deficit. 
Many origins have been proposed for the [\ion{C}{ii}] deficit. 
One possible origin is small dust destruction or charging, which reduces photoelectric yield (e.g., \citealp{Smith:2017aa}). 
If this is the case, the destruction of PAHs by strong radiation fields and large-scale shocks can naturally explain to reduce the [\ion{C}{ii}] emission during merging processes of galaxies. 

Finally, we comment out the limitation of our study. 
As described in Section 3.1, the photometric fluxes of the 21 galaxies with a companion, which are at early stages of merging processes, are used as the sum of the photometric flux of a primary galaxy and a companion galaxy due to the limitation of the low spatial resolution of IRAS. This can affect the interpretation of our results if the effect of the merging of galaxies is not the same in both galaxies. 
If the companion galaxy is significantly smaller than the primary galaxy, the summed SEDs are mainly reflected from the SED of the primary galaxy. If this is the case, we could not detect the PAH destruction of the companion galaxy from the summed SEDs. Furthermore, for the galaxies with a companion galaxy, we used the emission line ratios of primary galaxies due to the limitation of available spectroscopic data. 
Therefore, our result of the PAH destruction at early stages of merging processes and its origin may be uncertain. On the other hand, since the merger galaxies at the late stage are disturbed single component, it is certain that PAHs are destroyed at the late stages of merging processes of galaxies.

\section{Summary}
We have investigated the relation of $L_{\rm PAH}$, $L_{\rm TIR}$, $M_{\rm PAH}$, $M_{\rm VSG}$, $M_{\rm BG}$ and $M_{\rm star}$ with galaxy merger for 55 star-forming galaxies at redshift $z<0.2$ using the AKARI, WISE, IRAS, SDSS and HST data. We divided the galaxies into 35 merger galaxies and 20 non-merger galaxies based on $A$ which was calculated with the SDSS g-band for galaxies at $z<0.1$ and the HST F814W images for galaxies at $z>0.1$. We find that:  
\begin{enumerate}
  \item The fraction of merger galaxies clearly increases with $L_{\rm TIR}$, and the $L_{\rm PAH}/L_{\rm TIR}$ ratios of merger galaxies tend to be lower than those of non-merger galaxies, which results in the decrease of $L_{\rm PAH}/L_{\rm TIR}$ with increasing $L_{\rm TIR}$.
  \item Merger galaxies tend to show lower $M_{\rm PAH}/M_{\rm BG}$ than non-merger galaxies and there are no systematic differences in $M_{\rm VSG}/M_{\rm BG}$ and $M_{\rm BG}/M_{\rm star}$ between merger galaxies and non-merger galaxies. 
  \item Galaxies with very low $M_{\rm PAH}/M_{\rm BG}$ are clustered around the region occupied by the single ULIRGs in the $G$ and $M_{20}$ diagram, indicating that they are merger galaxies at late stages.
\end{enumerate}
These results suggest that PAHs are partly destroyed during merging processes of galaxies especially at the late stages. 

We consider three possible origins for the PAH destruction of star-forming galaxies; strong radiation fields, large-scale shocks by merging processes of galaxies, and both. 
To verify the above possibilities, we have investigated $M_{\rm PAH}/M_{\rm BG}$ variations in radiation field intensity strength $G_0$ and the emission line ratio of $[\ion{O}{i}]/{\rm H}\alpha$ which is a shock tracer for merger galaxies. We find that $M_{\rm PAH}/M_{\rm BG}$ decreases with increasing both $G_0$ and $[\ion{O}{i}]/{\rm H}\alpha$. 
Galaxies at a given $G_0$ tend to possess lower $M_{\rm PAH}/M_{\rm BG}$ with increasing $[\ion{O}{i}]/{\rm H}\alpha$. Galaxies at a given $[\ion{O}{i}]/{\rm H}\alpha$ tend to possess lower $M_{\rm PAH}/M_{\rm BG}$ with increasing $G_0$. 
Therefore, PAH destruction is likely to be caused by two processes; strong radiation fields and large-scale shocks during merging processes of galaxies. 

\section*{Acknowledgements}
We would like to thank the anonymous referee for valuable comments and suggestions that have improved the paper. 
This research is based on observations with AKARI, a JAXA project with the participation of ESA. 
We would like to thank all the members of AKARI project. 
% SDSS
Funding for SDSS-III has been provided by the Alfred P. Sloan Foundation, the Participating Institutions, the National Science Foundation, and the U.S. Department of Energy Office of Science. The SDSS-III web site is http://www.sdss3.org/.
SDSS-III is managed by the Astrophysical Research Consortium for the Participating Institutions of the SDSS-III Collaboration including the University of Arizona, the Brazilian Participation Group, Brookhaven National Laboratory, Carnegie Mellon University, University of Florida, the French Participation Group, the German Participation Group, Harvard University, the Instituto de Astrofisica de Canarias, the Michigan State/Notre Dame/JINA Participation Group, Johns Hopkins University, Lawrence Berkeley National Laboratory, Max Planck Institute for Astrophysics, Max Planck Institute for Extraterrestrial Physics, New Mexico State University, New York University, Ohio State University, Pennsylvania State University, University of Portsmouth, Princeton University, the Spanish Participation Group, University of Tokyo, University of Utah, Vanderbilt University, University of Virginia, University of Washington, and Yale University. 
% HLA
This research is based on observations made with the NASA/ESA Hubble Space Telescope, and obtained from the Hubble Legacy Archive, which is a collaboration between the Space Telescope Science Institute (STScI/NASA), the Space Telescope European Coordinating Facility (ST-ECF/ESA) and the Canadian Astronomy Data Centre (CADC/NRC/CSA). 
% WISE
This publication makes use of data products from the Wide-field Infrared Survey Explorer, which is a joint project of the University of California, Los Angeles, and the Jet Propulsion Laboratory/California Institute of Technology, funded by the National Aeronautics and Space Administration.

%%%%%%%%%%%%%%%%%%%%%%%%%%%%%%%%%%%%%%%%%%%%%%%%%%

%%%%%%%%%%%%%%%%% APPENDICES %%%%%%%%%%%%%%%%%%%%%

\appendix
\section{$G$ and $M_{20}$}
The Gini coefficient means inequality of a flux distribution and is defined as 
\begin{equation}
  G = \frac{1}{|\overline{X}|n(n-1)} \sum_i (2i-n-1) |X_i|,
\end{equation} 
where $X_i$ is the flux at the $i$-th brightest pixel values in the quasi-petrosian segmentation map, $\overline{X}$ is the average of $X_i$ and $n$ is the total number of pixels. The summation is done within the quasi-petrosian segmentation. 
For example, $G$ is 0 if all the pixels have the same value, while $G$ becomes closer to 1 if one pixel carries most of the total flux of a galaxy.

$M_{20}$ is a logarithmic fraction of the intensity momentum of the brightest region which includes 20\% flux of the total flux to the momentum of the total region of a galaxy. 
The $M_{20}$ is defined as
\begin{equation}
  M_{20} = \log_{10} \left(\frac{\sum_i M_i}{M_{\rm tot}}\right)~~{\rm while}~~\sum_i f_i < 0.2 f_{\rm tot}, 
\end{equation} 
where $M_i$ is the second-order momentum in the $i$-th pixel,  
\begin{equation}
  M_i = f_i  [(x_i-x_c)^2 + (y_i-y_c)^2], 
\end{equation}
where $x_i$ and $y_i$ are respectively the x-coordinate and y-coordinate of the $i$-th pixel, $x_c$ and $y_c$ are those of the central pixel of the galaxy, and $f_i$ and $f_{\rm tot}$ are the flux of the $i$-th pixel and the total flux respectively. $M_{\rm tot}$ is the summation of $M_i$ within the quasi-petrosian segmentation. The center of the galaxy~$(x_c, y_c)$~is determined to minimize the total momentum $M_{\rm tot}$. The summation of $M_{\rm i}$ is done within the brightest region which includes 20\% flux of the total flux.

% Don't change these lines
\bsp	% typesetting comment
\label{lastpage}

\begin{thebibliography}{}
\bibitem[Abraham et al.(1996)]{Abraham:1996jm} Abraham, R.~G., Tanvir, N.~R., Santiago, B.~X., et al.\ 1996, \mnras, 279, L47 
\bibitem[Abraham et al.(2007)]{Abraham:2007hy} Abraham, R.~G., Nair, P., McCarthy, P.~J., et al.\ 2007, \apj, 669, 184 
\bibitem[Ahn et al.(2012)]{Ahn:2012ih} Ahn, C.~P., Alexandroff, R., Allende Prieto, C., et al.\ 2012, \apjs, 203, 21 
\bibitem[Baldwin et al.(1981)]{Baldwin:1981aa} Baldwin, J.~A., Phillips, M.~M., \& Terlevich, R.\ 1981, \pasp, 93, 5 
\bibitem[Brinchmann et al.(2004)]{Brinchmann:2004aa} Brinchmann, J., Charlot, S., White, S.~D.~M., et al.\ 2004, \mnras, 351, 1151 
\bibitem[Capak et al.(2007)]{Capak:2007aa} Capak, P., Abraham, R.~G., Ellis, R.~S., et al.\ 2007, \apjs, 172, 284 
\bibitem[Chabrier(2003)]{Chabrier:2003aa} Chabrier, G.\ 2003, \pasp, 115, 763 
\bibitem[Chen et al.(2010)]{Chen:2010aa} Chen, Y., Lowenthal, J.~D., \& Yun, M.~S.\ 2010, \apj, 712, 1385 
\bibitem[Compi{\`e}gne et al.(2011)]{Compiegne:2011jf} Compi{\`e}gne, M., Verstraete, L., Jones, A., et al.\ 2011, \aap, 525, A103 
\bibitem[Conselice(2003)]{Conselice:2003ds} Conselice, C.~J.\ 2003, \apjs, 147, 1 
\bibitem[Conselice et al.(2008)]{Conselice:2008de} Conselice, C.~J., Rajgor, S., \& Myers, R.\ 2008, \mnras, 386, 909 
\bibitem[Cox et al.(2008)]{Cox:2008aa} Cox, T.~J., Jonsson, P., Somerville, R.~S., Primack, J.~R., \& Dekel, A.\ 2008, \mnras, 384, 386 
\bibitem[D{\'{\i}}az-Santos et al.(2013)]{Diaz-Santos:2013aa} D{\'{\i}}az-Santos, T., Armus, L., Charmandaris, V., et al.\ 2013, \apj, 774, 68 
\bibitem[D{\'{\i}}az-Santos et al.(2014)]{Diaz-Santos:2014aa} D{\'{\i}}az-Santos, T., Armus, L., Charmandaris, V., et al.\ 2014, \apjl, 788, L17 
\bibitem[Doi et al.(2015)]{Doi:2015cy} Doi, Y., Takita, S., Ootsubo, T., et al.\ 2015, \pasj, 67, 50 
\bibitem[Domingue et al.(2016)]{Domingue:2016aa} Domingue, D.~L., Cao, C., Xu, C.~K., et al.\ 2016, \apj, 829, 78 
\bibitem[Elbaz et al.(2011)]{Elbaz:2011ix} Elbaz, D., Dickinson, M., Hwang, H.~S., et al.\ 2011, \aap, 533, A119 
\bibitem[Engelbracht et al.(2005)]{Engelbracht:2005aa} Engelbracht, C.~W., Gordon, K.~D., Rieke, G.~H., et al.\ 2005, \apjl, 628, L29 
\bibitem[Eskew et al.(2012)]{Eskew:2012ft} Eskew, M., Zaritsky, D., \& Meidt, S.\ 2012, \aj, 143, 139 
\bibitem[Farage et al.(2010)]{Farage:2010aa} Farage, C.~L., McGregor, P.~J., Dopita, M.~A., \& Bicknell, G.~V.\ 2010, \apj, 724, 267 
\bibitem[Guillard et al.(2010)]{Guillard:2010aa} Guillard, P., Boulanger, F., Cluver, M.~E., et al.\ 2010, \aap, 518, A59 
\bibitem[Imanishi \& Dudley(2000)]{Imanishi:2000ev} Imanishi, M., \& Dudley, C.~C.\ 2000, \apj, 545, 701 
\bibitem[Imanishi et al.(2008)]{Imanishi:2008wj} Imanishi, M., Nakagawa, T., Ohyama, Y., et al.\ 2008, \pasj, 60, S489 
\bibitem[Imanishi et al.(2010)]{Imanishi:2010jw} Imanishi, M., Nakagawa, T., Shirahata, M., Ohyama, Y., \& Onaka, T.\ 2010, \apj, 721, 1233 
\bibitem[Ishihara et al.(2010)]{Ishihara:2010fi} Ishihara, D., Kaneda, H., Furuzawa, A., et al.\ 2010, \aap, 521, L61 
\bibitem[Ishihara et al.(2010)]{Ishihara:2010gd} Ishihara, D., Onaka, T., Kataza, H., et al.\ 2010, \aap, 514, A1 
\bibitem[Jones et al.(1996)]{Jones:1996bi} Jones, A.~P., Tielens, A.~G.~G.~M., \& Hollenbach, D.~J.\ 1996, \apj, 469, 740 
\bibitem[Jonsson et al.(2006)]{Jonsson:2006aa} Jonsson, P., Cox, T.~J., Primack, J.~R., \& Somerville, R.~S.\ 2006, \apj, 637, 255 
\bibitem[Kauffmann et al.(2003)]{Kauffmann:2003aa} Kauffmann, G., Heckman, T.~M., Tremonti, C., et al.\ 2003, \mnras, 346, 1055 
\bibitem[Kewley et al.(2001)]{Kewley:2001aa} Kewley, L.~J., Dopita, M.~A., Sutherland, R.~S., Heisler, C.~A., \& Trevena, J.\ 2001, \apj, 556, 121 
\bibitem[Kilerci Eser et al.(2014)]{KilerciEser:2014wv} Kilerci Eser, E., Goto, T., \& Doi, Y.\ 2014, \apj, 797, 54 
\bibitem[Kim et al.(2002)]{Kim:2002hc} Kim, D.-C., Veilleux, S., \& Sanders, D.~B.\ 2002, \apjs, 143, 277 
\bibitem[Landsman(1993)]{Landsman:1993aa} Landsman, W.~B.\ 1993, Astronomical Data Analysis Software and Systems II, 52, 246 
\bibitem[Larson et al.(2016)]{Larson:2016aa} Larson, K.~L., Sanders, D.~B., Barnes, J.~E., et al.\ 2016, \apj, 825, 128 
\bibitem[Law et al.(2012)]{Law:2012hf} Law, D.~R., Steidel, C.~C., Shapley, A.~E., et al.\ 2012, \apj, 745, 85 
\bibitem[Lequeux(2005)]{Lequeux:2005aa} Lequeux, J.\ 2005, The interstellar medium, Translation from the French language edition of: Le Milieu Interstellaire by James Lequeux, EDP Sciences, 2003 Edited by J.~Lequeux.~ Astronomy and astrophysics library, Berlin: Springer, 2005,  
\bibitem[Li \& Draine(2001)]{Li:2001aa} Li, A., \& Draine, B.~T.\ 2001, \apjl, 550, L213 
\bibitem[Lotz et al.(2004)]{Lotz:2004gj} Lotz, J.~M., Primack, J., \& Madau, P.\ 2004, \aj, 128, 163 
\bibitem[Lotz et al.(2008)]{Lotz:2008wg} Lotz, J.~M., Jonsson, P., Cox, T.~J., \& Primack, J.~R.\ 2008, \mnras, 391, 1137 
\bibitem[Lotz et al.(2010)]{Lotz:2010hf} Lotz, J.~M., Jonsson, P., Cox, T.~J., \& Primack, J.~R.\ 2010, \mnras, 404, 575 
\bibitem[Lotz et al.(2011)]{Lotz:2011bu} Lotz, J.~M., Jonsson, P., Cox, T.~J., et al.\ 2011, \apj, 742, 103 
\bibitem[Luhman et al.(2003)]{Luhman:2003aa} Luhman, M.~L., Satyapal, S., Fischer, J., et al.\ 2003, \apj, 594, 758 
\bibitem[Micelotta et al.(2010)]{Micelotta:2010dl} Micelotta, E.~R., Jones, A.~P., \& Tielens, A.~G.~G.~M.\ 2010, \aap, 510, A36 
\bibitem[Mihos \& Hernquist(1996)]{Mihos:1996aa} Mihos, J.~C., \& Hernquist, L.\ 1996, \apj, 464, 641 
\bibitem[Monreal-Ibero et al.(2006)]{MonrealIbero:2006fi} Monreal-Ibero, A., Arribas, S., \& Colina, L.\ 2006, \apj, 637, 138 
\bibitem[Monreal-Ibero et al.(2010)]{MonrealIbero:2010gz} Monreal-Ibero, A., Arribas, S., Colina, L., et al.\ 2010, \aap, 517, A28 
\bibitem[Moorwood(1986)]{Moorwood:1986wa} Moorwood, A.~F.~M.\ 1986, \aap, 166, 4 
\bibitem[Moshir et al.(1990)]{Moshir:1990vn} Moshir, M., et al.\ 1990, IRAS Faint Source Catalogue, version 2.0
\bibitem[Murakami et al.(2007)]{Murakami:2007bs} Murakami, H., Baba, H., Barthel, P., et al.\ 2007, \pasj, 59, S369 
\bibitem[Murata et al.(2014)]{Murata:2014hy} Murata, K., Matsuhara, H., Inami, H., et al.\ 2014, \aap, 566, A136 
\bibitem[Neugebauer et al.(1984)]{Neugebauer:1984ch} Neugebauer, G., Habing, H.~J., van Duinen, R., et al.\ 1984, \apjl, 278, L1 
\bibitem[Nordon et al.(2012)]{Nordon:2012io} Nordon, R., Lutz, D., Genzel, R., et al.\ 2012, \apj, 745, 182 
\bibitem[O'Halloran et al.(2006)]{OHalloran:2006eu} O'Halloran, B., Satyapal, S., \& Dudik, R.~P.\ 2006, \apj, 641, 795 
\bibitem[Ohyama et al.(2007)]{Ohyama:2007gm} Ohyama, Y., Onaka, T., Matsuhara, H., et al.\ 2007, \pasj, 59, S411 
\bibitem[Onaka et al.(2007)]{Onaka:2007pj} Onaka, T., Matsuhara, H., Wada, T., et al.\ 2007, \pasj, 59, S401 
\bibitem[Oyabu et al.(2011)]{Oyabu:2011kh} Oyabu, S., Ishihara, D., Malkan, M., et al.\ 2011, \aap, 529, A122 
\bibitem[Peeters et al.(2004)]{Peeters:2004ga} Peeters, E., Spoon, H.~W.~W., \& Tielens, A.~G.~G.~M.\ 2004, \apj, 613, 986 
\bibitem[Plante \& Sauvage(2002)]{Plante:2002jk} Plante, S., \& Sauvage, M.\ 2002, \aj, 124, 1995 
\bibitem[Psychogyios et al.(2016)]{Psychogyios:2016aa} Psychogyios, A., Charmandaris, V., Diaz-Santos, T., et al.\ 2016, \aap, 591, A1 
\bibitem[Rich et al.(2011)]{Rich:2011is} Rich, J.~A., Kewley, L.~J., \& Dopita, M.~A.\ 2011, \apj, 734, 87 
\bibitem[Rich et al.(2014)]{Rich:2014ib} Rich, J.~A., Kewley, L.~J., \& Dopita, M.~A.\ 2014, \apjl, 781, L12 
\bibitem[Rich et al.(2015)]{Rich:2015kf} Rich, J.~A., Kewley, L.~J., \& Dopita, M.~A.\ 2015, \apjs, 221, 28 
\bibitem[Risaliti et al.(2006)]{Risaliti:2006bm} Risaliti, G., Maiolino, R., Marconi, A., et al.\ 2006, \mnras, 365, 303 
\bibitem[Sanders \& Mirabel(1996)]{Sanders:1996cd} Sanders, D.~B., \& Mirabel, I.~F.\ 1996, \araa, 34, 749 
\bibitem[Schade et al.(1995)]{Schade:1995ie} Schade, D., Lilly, S.~J., Crampton, D., et al.\ 1995, \apjl, 451, L1 
\bibitem[Smith et al.(2007)]{Smith:2007aa} Smith, J.~D.~T., Draine, B.~T., Dale, D.~A., et al.\ 2007, \apj, 656, 770 
\bibitem[Smith et al.(2017)]{Smith:2017aa} Smith, J.~D.~T., Croxall, K., Draine, B., et al.\ 2017, \apj, 834, 5 
\bibitem[Snyder et al.(2015)]{Snyder:2015aa} Snyder, G.~F., Lotz, J., Moody, C., et al.\ 2015, \mnras, 451, 4290 
\bibitem[Springel et al.(2005)]{Springel:2005aa} Springel, V., Di Matteo, T., \& Hernquist, L.\ 2005, \mnras, 361, 776 
\bibitem[Stierwalt et al.(2013)]{Stierwalt:2013aa} Stierwalt, S., Armus, L., Surace, J.~A., et al.\ 2013, \apjs, 206, 1 
\bibitem[Stierwalt et al.(2014)]{Stierwalt:2014aa} Stierwalt, S., Armus, L., Charmandaris, V., et al.\ 2014, \apj, 790, 124 
% \bibitem[Springel et al.(2005)]{Springel:2005aa} Springel, V., Di Matteo, T., \& Hernquist, L.\ 2005, \mnras, 361, 776 
\bibitem[Tasca et al.(2009)]{Tasca2009aa} Tasca, L.~A.~M., Kneib, J.-P., Iovino, A., et al.\ 2009, \aap, 503, 379 
\bibitem[Teyssier et al.(2010)]{Teyssier2010aa} Teyssier, R., Chapon, D., \& Bournaud, F.\ 2010, \apjl, 720, L149 
\bibitem[Tielens(2008)]{Tielens:2008fx} Tielens, A.~G.~G.~M.\ 2008, \araa, 46, 289 
\bibitem[Tremonti et al.(2004)]{Tremonti:2004aa} Tremonti, C.~A., Heckman, T.~M., Kauffmann, G., et al.\ 2004, \apj, 613, 898 
\bibitem[Veilleux et al.(1999)]{Veilleux:1999aa} Veilleux, S., Kim, D.-C., \& Sanders, D.~B.\ 1999, \apj, 522, 113 
\bibitem[Veilleux et al.(2002)]{Veilleux:2002dj} Veilleux, S., Kim, D.-C., \& Sanders, D.~B.\ 2002, \apjs, 143, 315 
\bibitem[Wang et al.(2006)]{Wang:2006hu} Wang, J.~L., Xia, X.~Y., Mao, S., et al.\ 2006, \apj, 649, 722 
\bibitem[Wright et al.(2010)]{Wright:2010in} Wright, E.~L., Eisenhardt, P.~R.~M., Mainzer, A.~K., et al.\ 2010, \aj, 140, 1868-1881 
\bibitem[Yamada et al.(2013)]{Yamada:2013ca} Yamada, R., Oyabu, S., Kaneda, H., et al.\ 2013, \pasj, 65,  
\bibitem[Zamojski et al.(2007)]{Zamojski:2007gn} Zamojski, M.~A., Schiminovich, D., Rich, R.~M., et al.\ 2007, \apjs, 172, 468 
\end{thebibliography}
\end{document}